\documentclass[10pt,aps,prd,twocolumn,amsmath,amssymb,floatfix,superscriptaddress,nofootinbib,preprintnumbers]{revtex4-1}

\usepackage{amsfonts,amssymb,amsmath,graphicx,color,bm,epsfig}
\usepackage[normalem]{ulem}
\usepackage{multirow}

\definecolor{ultramarine}{rgb}{0.07, 0.04, 0.56}
\definecolor{cadmiumgreen}{rgb}{0.0, 0.42, 0.24}
\definecolor{indigo(dye)}{rgb}{0.0, 0.25, 0.42}
\usepackage[linktocpage=true]{hyperref}
\hypersetup{
colorlinks=true,
citecolor=ultramarine,
linkcolor=cadmiumgreen,
urlcolor=indigo(dye),
}

\newcommand{\be}{\begin{equation}}  
\newcommand{\ee}{\end{equation}}

\newcommand{\Mpl}{M_{\rm Pl}}
\renewcommand{\d}{\delta}
\newcommand{\e}{\epsilon_H}

\newcommand{\C}{\mathcal{C}}
\renewcommand{\O}{\mathcal{O}}
\newcommand{\Ha}{\mathcal{H}}
\renewcommand{\L}{\mathcal{L}}
\newcommand{\pa}{\partial}

\newcommand*\diff{\mathop{}\!d}
\newcommand*\diffcubed{\mathop{}\!d^3}
\newcommand*\difffour{\mathop{}\!d^4}

\newcommand*\re[1]{\mathop{}\!\mathrm{Re}\left[#1 \right]}
\newcommand*\im[1]{\mathop{}\!\mathrm{Im}\left[#1 \right]}
\newcommand\numberthis{\addtocounter{equation}{1}\tag{\theequation}}

\begin{document}

\title[EFT NG]{Scalar bispectrum beyond slow-roll in the unified EFT of inflation}

\author{Samuel Passaglia}
\email{passaglia@uchicago.edu}
\affiliation{Kavli Institute for Cosmological Physics, Department of Astronomy \& Astrophysics, 
Enrico Fermi Institute, University of Chicago, Chicago, IL 60637}

\author{Wayne Hu}
\affiliation{Kavli Institute for Cosmological Physics, Department of Astronomy \& Astrophysics, 
Enrico Fermi Institute, University of Chicago, Chicago, IL 60637}

\label{firstpage}

\begin{abstract}
We present a complete formulation of the scalar bispectrum in the unified effective field theory (EFT) of inflation, which includes the Horndeski and beyond-Horndeski Gleyzes-Langlois-Piazza-Vernizzi classes, in terms of a set of
 simple one-dimensional integrals. These generalized slow-roll expressions  remain valid even when slow-roll is transiently violated and encompass all configurations of the bispectrum. We show analytically that our expressions explicitly preserve the squeezed-limit consistency relation beyond slow-roll. As an example application of our results, we compute the scalar bispectrum in a model in which potential-driven G-inflation at early times transitions to chaotic inflation at late times, showing that our expressions accurately track the bispectrum when slow-roll is violated and conventional slow-roll approximations fail.
\end{abstract}

\date{\today}

\maketitle 

\section{Introduction}
	\label{sec:Intro}

	Superhorizon curvature perturbations in the cosmic microwave background imply a period of exponential expansion before nucleosynthesis \cite{Ade:2015xua}. Moving beyond this simple phenomenological picture to understand the physics driving the exponential expansion requires finding other observable signatures of inflation.

	The non-Gaussianity of the primordial fluctuations is a powerful such probe \cite{Chen:2006nt,Maldacena:2002vr,Martin:2012pe,Mizuno:2010ag,Ribeiro:2011ax,Chen:2006xjb,Creminelli:2011rh,Seery:2005gb,Senatore:2009gt,Weinberg:2005vy,Meerburg:2016zdz}. That the bispectrum is as small	as current upper limits require is not {\it a priori} given in well-motivated constructions of inflation (see, e.g., Ref.~\cite{Baumann:2011su}) and already significantly constrains the inflaton sound speed and the presence of features during inflation \cite{Ade:2015lrj}.

	Testing the single-field hypothesis requires making these statements precise by studying the general predictions of single-field inflation as well as the specific predictions of individual single-field models. The effective field theory (EFT) of inflation is a powerful framework in which to do so \cite{Creminelli:2006xe,Cheung:2007st,Cheung:2007sv}. Refs.~\cite{Gleyzes:2013ooa,Kase:2014cwa,Gleyzes:2014dya,Gleyzes:2014qga,Motohashi:2017gqb} recently extended the EFT of inflation to include a complete set of  ADM operators  for which the lapse and shift remain nondynamical.   Scalar and tensor power spectrum observables of this `unified' EFT of inflation were studied beyond slow-roll for terms that 
	  lead to metric perturbations with a standard dispersion
	relation
	in Ref.~\cite{Motohashi:2017gqb}.

	In this paper, we extend Ref.~\cite{Motohashi:2017gqb} to study the scalar bispectrum beyond
	the slow-roll approximation in the unified EFT of inflation using techniques
	developed in Ref.~\cite{Adshead:2013zfa}. Because inflation need not obey this approximation during its entire course, there is a rich range of phenomenological possibilities in the bispectrum of single field inflation \cite{Adshead:2011jq,Flauger:2009ab,Chen:2008wn,Chen:2010bka,Ribeiro:2012ar,Burrage:2011hd}.

	We structure our results such that existing models, such as those in the Horndeski \cite{Horndeski:1974wa} or {Gleyzes-Langlois-Piazza-Vernizzi (GLPV)} classes \cite{Gleyzes:2014qga}, can be straightforwardly plugged into our expressions, and we write our results in such a way as to manifestly preserve the model-independent consistency relation between the squeezed-limit of the bispectrum and the slope of the power spectrum in and beyond slow-roll.
	
	This paper is organized as follows. In \S\ref{sec:EFT}, we construct the unified EFT of inflation, derive the corresponding cubic action for scalar metric perturbations, and study the structure of the cubic action in the Horndeski and GLPV subclasses. The explicit EFT coefficients which make up the cubic action are provided in Appendix~\ref{app:Fs}. In \S\ref{sec:GSR}, we construct an integral formulation of the bispectrum to first order in the generalized slow-roll (GSR) formalism and show that the consistency relation between the power spectrum and the squeezed limit of the bispectrum is explicitly preserved beyond slow-roll for sharp features. The full set of integral sources, windows, and  configuration weights  which make up our bispectrum formulation are provided in Appendix~\ref{app:EFTIntegrals}. In \S\ref{sec:GInflation}, we provide an example application of our bispectrum formulation by explicitly computing the bispectrum in a specific inflationary model, transient G-inflation
	\cite{Ramirez:2018dxe}.  We conclude in \S\ref{sec:Discussion} by discussing our results in the context of related works. 

	We use the $(-+++)$ metric signature and set $\Mpl = 1$ throughout.

\section{Unified EFT of Inflation}
	\label{sec:EFT}

	In this section we derive the cubic action for scalar metric perturbations in the unified EFT of inflation. We begin in \S\ref{subsec:EFTLagrangian} by reviewing and generalizing the construction of the Lagrangian of the EFT of inflation, which we then expand to cubic order in scalar metric perturbations in \S\ref{subsec:EFTCubicAction}. We rewrite this action to make the squeezed-limit consistency relation manifest in \S\ref{subsec:CubicActionConsistency}. Finally, we study the structure of the EFT in the Horndeski and beyond-Horndeski GLPV limits in \S\ref{subsec:HorndeskiGLPV}.

	In general, we find that the cubic action for scalar perturbations can be written in terms of ten operators and manifestly leads to the squeezed-limit consistency relation during slow-roll. In the Horndeski and GLPV subclasses, six of the ten operators are present.

	\subsection{Lagrangian}
		\label{subsec:EFTLagrangian}

		The unified EFT of inflation was presented in Ref.~\cite{Motohashi:2017gqb} with the complete set of quadratic operators that contribute  to theories where the metric perturbations obey a second-order equation in both
		time and space and temporal components of the metric remain nondynamical. These restrictions ensure that the power spectra of scalar and tensor metric fluctuations 
		obey their usual form. We summarize here some of the essential features of that construction while extending it to include the complete set of cubic operators that contribute to the bispectrum.

		In the EFT construction, we seek the most general form for the action that is consistent with  unbroken spatial diffeomorphisms and	a preferred temporal coordinate that represents the ``clock" during inflation.  Using this preferred slicing, we decompose the metric into its $3+1$ ADM form
		\begin{equation}
		\label{eq:ADMMetric}
		\diff{s}^2 = -N^2 \diff{t}^2 + h_{ij} (\diff{x}^i + N^i \diff{t})(\diff{x}^j + N^j \diff{t}),
		\end{equation}
		with the lapse $N$, the shift $N^i$, and the spatial metric $h_{ij}$.  

		This metric and a unit timelike vector $n_\mu$ orthogonal to constant $t$ surfaces define the spatial tensors  that compose the EFT action. We construct an action invariant under spatial diffeomorphisms out of a general scalar function of these quantities
		\begin{equation}
		\label{eq:EFTAction} 
		S = \int \difffour{x} N \sqrt{h} \, L(N, {K^i}_j, {R^i}_j, t),
		\end{equation}
		in which $K_{\mu \nu} = n_{\mu;\nu} + n_\nu  n_{\mu;\beta} n^\beta$ is the extrinsic curvature, $R_{ij}$ is the three-dimensional Ricci tensor with trace $R ={R^i}_i$, and $h$ is the determinant of the three-dimensional metric $h_{ij}$. Semicolons here and throughout denote covariant derivatives with respect to the metric $g_{\mu \nu}$. Latin indices denote spatial coordinates, which are raised and lowered using $h_{ij}$. We use the shorthand summation convention
		\begin{equation}
		S_{i \ldots j}T_{i \ldots j}   \equiv \d^{i i'} \ldots \d^{j j'} S_{i \ldots j} T_{i' \ldots j'}, 
		\end{equation}
		for any two spatial tensors $S$ and $T$.

		We have not allowed additional spatial derivatives in  Eq.~\eqref{eq:EFTAction} since they lead to equations of motion that are	beyond second order in spatial derivatives.  Thus we do not encompass the spatially covariant gravity \cite{Gao:2014soa,Gao:2014fra} or the Ho\v{r}ava-Lifshitz theories \cite{Horava:2009uw,Blas:2009yd,Blas:2010hb}. We 
		have also not allowed the lapse or shift to be dynamical, and thus we do not encompass the full set of degenerate higher-order scalar tensor  (DHOST)\cite{BenAchour:2016fzp,Langlois:2017mxy} theories.

		Next we perturb the action \eqref{eq:EFTAction} around a spatially flat FLRW background, 
		\begin{equation}
		\left[N\right] = 1, \quad \left[N^i\right] = 0, \quad \left[h_{ij}\right] = a^2 \d_{ij},
		\end{equation}
		on which the extrinsic and intrinsic curvature are		
		\begin{equation}
		\left[{K^i}_j\right] = H {\d^i}_j, \quad \left[{R^i}_j\right] = 0,
		\end{equation}
		with $H \equiv d{\ln{a}}/dt$. Here and below the notation $\left[ \ldots \right]$ denotes evaluation on the background.

		In order to keep all terms that are at most cubic in metric perturbations, we expand the Lagrangian to cubic order in the ADM variables around the background. We define the Taylor coefficients
		\begin{align*}
		\label{eq:lagrangianpartials}
		\left[L\right] & =  \C, \numberthis\\
		\left[\frac{\partial L}{\partial X^i_{\,j}} \right] &=  \C_X {\delta_i^{j}}, \\ 
		\left[\frac{\partial^2 L}{\partial X^i_{\,j} \partial Y^k_{\hphantom{k}l}}\right] &=
		 \C_{XY} \delta_i^{\,j}  \delta_k^{\,l} + \frac{ \C_{\bar{X}\bar{Y}}}{2} (\delta^l_i \delta^j_k + \delta_{ik} \delta^{jl}), \\ 
		\left[\frac{\partial^3 L}{\partial X^i_{\,j} \partial Y^k_{\hphantom{k}l} \partial Z^m_{\,\,\, n}}\right] &= \C_{XYZ} \delta^j_i \delta^l_k \delta^n_m 
		\\&\quad+ \frac{\C_{\bar{X} \bar{Y} Z}}{2}  \delta^n_m (\delta^l_i \delta^j_k + \delta_{ik} \delta^{jl}) 
		\\&\quad+ \frac{\C_{X \bar{Y} \bar{Z}}}{2}  \delta^j_i (\delta^l_m \delta^n_k + \delta_{mk} \delta^{nl}) 
		\\&\quad+ \frac{\C_{\bar{X} Y \bar{Z}}}{2}  \delta^l_k (\delta^n_i \delta^j_m + \delta_{im} \delta^{jn})  
		\\&\quad+ \frac{\C_{\bar{X}\bar{Y}\bar{Z}}}{8} (\delta^j_k \delta^l_m \delta^n_i + \delta^j_k \delta^{l n} \delta_{m i} 
		\\&\quad+ \delta^{j l} \delta^n_k \delta_{m i} + \delta^{j l} \delta_{k m} \delta^n_i + \delta_{k m}  \delta^l_i \delta^{n j} 
		\\&\quad+ \delta^n_k \delta^j_m \delta^l_i  + \delta_{i k} \delta^{n j} \delta^l_m + \delta_{i k} \delta^j_m \delta^{n l}  ), 
		\end{align*}
		where $X, Y, Z \in \{N, K, R\}$ and the index structure is determined by the symmetry of the background. We treat scalars and traces with the same notation, so that the tensor ${N^i}_j = (N/3)\delta^i_j$. Thus $\C_{\bar{N}\bar{X}Y} = \C_{\bar{N}\bar{X}\bar{Y}}  = 0$ for any $X, Y$.  Otherwise, these coefficients are arbitrary functions of time which are invariant under subscript permutation in the EFT;
		they take different concrete forms in different specific inflationary models. Notationally, our $\C_{\bar{X}\bar{Y}}$ is equal to the $\tilde\C_{XY}$ of Ref.~\cite{Motohashi:2017gqb}. Up to cubic order we can write
		\begin{align*}
		\label{eq:cubicADMLagrangian}
			L =&\ \frac{1}{3!} \sum_{X,Y,Z} \left(\C_{XYZ} \d X \d Y \d Z + \C_{X\bar{Y}\bar{Z}} \d X \d {Y^i}_j \d {Z^j}_i \right. \\ 
			   &+ \C_{\bar{X}Y\bar{Z}} \d {X^i}_j \d Y \d {Z^j}_i  + \C_{\bar{X}\bar{Y}Z} \d {X^i}_j \d {Y^j}_i \d Z \\ 
			   &+ \left. \C_{\bar{X}\bar{Y}\bar{Z}} \d {X^i}_j \d {Y^j}_k \d {Z^k}_i \right) \\
			   &+ \frac{1}{2} \sum_{Y,Z} \left(\C_{YZ} \d Y \d Z + \C_{\bar{Y}\bar{Z}} \d {Y^i}_j \d {Z^j}_i \right) \\
			   &+ \C_N \d N + \C_R \d R + \frac{\C_N}{N} - \C_N,
			\numberthis
		\end{align*}
		with the sums running through all variable permutations with replacement. We have followed Ref.~\cite{Motohashi:2017gqb} in using integration by parts to eliminate the linear $\d K$ term up to a total derivative term as well as in using the background equation of motion to simplify some of the terms which are constant or linear in geometric quantities. 
		
		Finally, to ensure only second-order spatial derivatives in the equation of motion of perturbations we impose
		\begin{align*}
			\label{eq:QuadraticInheritance}
			\C_{\bar{K} \bar{K}} &= - \C_{KK},\\
			\C_{\bar{K}\bar{R}} &= - 2 \C_{KR},\\
			\C_{\bar{R}\bar{R}} &= -\frac{8}{3} \C_{RR}.
			\numberthis
		\end{align*}	
		This includes the Horndeski and GLPV classes.

	\subsection{Scalar perturbations}
		\label{subsec:EFTCubicAction}

		We now restrict our attention to scalar metric perturbations and derive the quadratic and cubic actions for their dynamical
		field, the curvature perturbation. For scalar perturbations the ADM metric \eqref{eq:ADMMetric} takes the form		\begin{equation}
		N = 1+\d N, \quad N_i = \pa_i \psi, \quad h_{ij} = a^2e^{2\zeta} \d_{ij},
		\end{equation}
		where we have fixed the residual gauge freedom associated with spatial diffeomorphism invariance by taking a diagonal
		form for $h_{ij}$ \cite{Motohashi:2016prk}.   We call this choice unitary gauge.

		In unitary gauge, the perturbed geometric quantities are
		\begin{align*}
		\label{eq:ADMblocks}
		\d {K^i}_j =&\ \frac{1}{1+\d N}\Bigl[{\d^i}_j \left(\dot{\zeta}-H \d N\right)+ a^{-2} e^{-2\zeta} \left(\d^{ik} \pa_{k} \zeta \pa_{j} \psi \right. \\
		&+\d^{ik} \pa_{j} \zeta \pa_{k} \psi - \left. \d^{ik} \pa_k \pa_j \psi - {\d^i}_j \d^{ab} \pa_a \zeta \pa_b \psi\right) \Bigr],\\
		{\d R^i}_j =& -a^{-2} e^{-2\zeta} \left[\d^{ik} \pa_k \pa_j \zeta + {\d^{i}}_j \pa^2 \zeta + {\d^i}_j (\pa \zeta)^2 \right. \\ 
					&- \left. \d^{ik}\pa_k \zeta \pa_j \zeta  \right].
		\numberthis
		\end{align*}
		Here and throughout, $(\pa \zeta)^2 \equiv \d^{a b} \pa_a \zeta \pa_b \zeta$ and $\pa^2 \zeta \equiv \d^{a b} \pa_a \pa_b \zeta$.
		Variation of the quadratic action with respect to the lapse and shift yields the Hamiltonian and momentum constraints
		\begin{equation}
		\label{eq:constraints}
		\d N = D_1 \dot{\zeta}, \quad \psi = D_2 \zeta + a^2 D_3 \chi,
		\end{equation}
		where $\chi$ is an auxiliary variable satisfying  $\pa^{2} \chi = \dot{\zeta}$ and the parameters $D_1$, $D_2$, and $D_3$ are
		\begin{align*}
		D_1 &= \frac{2 \C_{KK}}{2 H \C_{KK} - \C_{NK}}, \\
		D_2 &= \frac{4(\C_{NR} + \C_{R}-H \C_{KR})}{2 H \C_{KK} - \C_{NK}}, \\
		D_3 &= \frac{3 \C_{NK}^2 - 2 \C_{KK} (2 \C_{N} + \C_{NN})}{(2 H \C_{KK} - \C_{NK})^2}.
		\label{eq:Dn}
		\numberthis
		\end{align*}

		Since we are interested in the action to cubic order in perturbations, the lapse and shift should a priori be expanded beyond linear order. However, direct computation shows that the $\O{(\zeta^2)}$ lapse and shift parameters do not contribute to the cubic action. This is an example of the general result that the $\O{(\zeta^2)}$ lapse and shift parameters multiply the order $\O{(\zeta)}$ constraint equations and therefore do not contribute to the cubic action \cite{Maldacena:2002vr,Chen:2006nt,Pajer:2016ieg}.		
		
		After eliminating the lapse and shift, the quadratic action for the curvature $\zeta$ becomes
		\begin{equation}
		S_2 = \int \difffour{x} \ a^3 Q \left[ \dot{\zeta}^2 - \frac{c_s^2}{a^2} \left(\pa \zeta\right)^2 \right],
		\end{equation}
		in which $Q$ and $c_s^2$ are
		\begin{align*}
		Q &= \frac{\C_{KK} \left(2 \C_{KK} \left(2 \C_{N} + \C_{NN}\right)- 3 \C_{NK}^2\right)}{\left(2 H \C_{KK}-\C_{NK}\right)^2}, \numberthis\\
		c_s^2 &= \frac{2}{a Q} \biggl[\frac{\diff{}}{\diff{t}} \left(a \frac{2 \C_{KK} (\C_{NR} + \C_R)-\C_{KR} \C_{NK}}{2H \C_{KK}-\C_{NK}}\right) -  a \C_R\biggr].
		\end{align*}
		In terms of the $b_s$ parameter defined in Ref.~\cite{Motohashi:2017gqb}, $Q \equiv \e b_s / c_s^2$. Here and throughout, $\e \equiv -\dot{H}/H^2$.
		The quadratic action provides the linearized equation of motion 
		\begin{equation}
		\label{eq:eom}
		\pa^2 \zeta = \frac{1}{a Q c_s^2} \frac{\diff{}}{\diff{t}} (a^3 Q \dot{\zeta}).
		\end{equation}

		We now plug in the perturbed geometric quantities \eqref{eq:ADMblocks} into the action \eqref{eq:EFTAction} with the Lagrangian \eqref{eq:cubicADMLagrangian}, eliminating the lapse and shift using the constraint equations \eqref{eq:constraints} and retaining terms up to cubic order in $\zeta$. We can also simplify the resulting action using integration by parts. Spatial boundary terms will not contribute to the in-in bispectrum, by momentum conservation, and will be omitted. Temporal boundary terms can contribute significantly and therefore must be retained \cite{Arroja:2011yj,Rigopoulos:2011eq}. 
				
		Finally, we can also use the linear equation of motion \eqref{eq:eom} to eliminate $\ddot\zeta$-type terms \cite{Burrage:2011hd,RenauxPetel:2011sb,Seery:2005gb}. The resulting cubic action is
		\begin{align*}
		\label{eq:cubiclagrangiandirect}
			S_3 =& \ S_3^{\text{Boundary}} + \int \diffcubed{x} \diff{t} \Bigl[a^3 F_1 \zeta  \dot\zeta {}^2 
			 + a F_2 \zeta  \left(\pa \zeta \right)^2 \numberthis \\
			 &+ a^3 \frac{F_3}{{H}} \dot\zeta {}^3
			 + a^3 F_4 \dot\zeta  \pa_a\zeta  \pa_a\chi 
			 + a^3 F_5 \pa^2\zeta  \left(\pa \chi \right)^2 \\
			 &+ \frac{F_6}{{H^3} a} \dot\zeta  \pa^2\zeta  \pa^2\zeta
			 + \frac{F_7}{{H^4} a^3} \left(\pa_a\pa_b\zeta \right)^2 \pa^2\zeta\\
			 &+ \frac{F_8}{{H^4} a^3} \pa^2\zeta  \pa^2\zeta  \pa^2\zeta 
			 + \frac{F_9}{{H^3} a} \pa^2\zeta \left(\pa_a\pa_b\zeta\right) \left(\pa_a\pa_b\chi\right) \Bigr],
		\end{align*}
		in which $F_1$ through $F_9$ are {dimensionless} time-dependent functions presented in Appendix~\ref{app:Fs}. The temporal boundary terms are
		\begin{align*}
		\label{eq:boundaryaction}
			S_3^{\text{Boundary}} =& \int \diffcubed{x} \diff{t} \frac{\diff{}}{\diff{t}} \left[ a^3 G_1 \dot\zeta {}^3 
			 + a^3 G_2 \zeta  \dot\zeta {}^2 \right. \\
			 &+ a G_3 \zeta  \left(\pa \zeta \right)^2 
			 + a^3 G_4 \dot\zeta  \pa_a\zeta  \pa_a\chi  \\
			 &+ a G_5 \dot\zeta  \left(\pa \zeta \right)^2 
			 + \frac{G_6}{a} \left(\pa \zeta \right)^2 \pa^2\zeta\\
			 &+ a^3 G_7 \pa^2\zeta  \left(\pa \chi \right)^2
			 + a G_8 \pa_a\zeta  \pa_b\zeta  \pa_a\pa_b\chi\\  
			 &+ \left. a^3 G_9 \dot\zeta  \left(\pa_a\pa_b\chi \right)^2  \right],
	 	\numberthis
		\end{align*}
		in which $G_1$ through $G_9$ are time-dependent functions. 

		The $G_3$ and $G_6$ terms contain no time-derivatives of the fields and therefore do not contribute to
		bispectrum in the in-in formalism regardless of the behavior of their coefficients \cite{Burrage:2011hd,Adshead:2011bw}.

 		The remaining terms are suppressed relative to the usual $a^3 \zeta^2\dot\zeta$ boundary operator, which shall appear later in our construction, by the presence either of spatial derivatives, which yield relative factors of $k/aH \ll 1$, or by the presence of additional factors of $\dot{\zeta}$, which is suppressed outside the horizon. Therefore none of these terms contribute unless $G_n$ grows sufficiently quickly, so long as the boundary is taken when all modes are outside the horizon.

		We restrict our attention to scenarios which satisfy these mild conditions on the EFT parameters and therefore we hereafter discard $S_3^{\text{Boundary}}$ entirely.

	\subsection{Cubic action and consistency relation}
		\label{subsec:CubicActionConsistency}
	
		We can use to our advantage our ability to reorganize the cubic action using integration by parts and the equation of motion for $\zeta$ derived from the
		quadratic action. In particular, it is well known that in inflation with a single dynamical degree of freedom and a curvature perturbation which remains constant outside the horizon, the bispectrum in the squeezed limit should satisfy the consistency relation \cite{Cheung:2007sv,Creminelli:2004yq,Maldacena:2002vr,Creminelli:2011rh}
		\begin{equation}
		\lim_{k_S\rightarrow0} \frac{B_\zeta(k_S, k_L, k_L)}{P_\zeta(k_S) P_\zeta(k_L)} = -\frac{d \ln{ \Delta_\zeta^2(k_L)}}{d \ln{k_L}},
		\label{eq:ConsistencyRelation}
 		\end{equation}
		where $B_\zeta$ denotes the curvature bispectrum (see \S\ref{sec:GSR} for notation).  Here the power spectrum $P_\zeta$ is related to the dimensionless power spectrum $\Delta_\zeta^2$ by
		\begin{equation}
		\frac{k^3}{2 \pi^2} P_\zeta \equiv \Delta_\zeta^2 
		\simeq \frac{H^2}{8\pi^2 Q c_s^3},
		\end{equation}
		where here and below $\simeq$ denotes a slow-roll relation.
		In the slow-roll approximation, the local slope of the power spectrum is nearly constant and is called the tilt 
		\begin{equation}
		\label{eq:tilt}
			\frac{d \ln \Delta_\zeta^2 }{d \ln k}  \simeq 	n_s - 1=  (-2 \e - q - 3 \sigma),
		\end{equation}
		where $q \equiv \dot{Q}/(HQ)$, $\sigma \equiv \dot{c}_s/(H c_s)$.		

		We expect the consistency relation to hold here, but at first glance -- or, in the language of \S\ref{sec:GSR}, when plugging in zeroth-order modefunctions --  the squeezed-contributing interactions $\zeta \dot{\zeta}^2$ and $\zeta (\pa \zeta)^2$ with their sources $F_1$ and $F_2$ are not obviously related to the tilt \eqref{eq:tilt}. We can rewrite these terms in such a way as to make the consistency relation manifest by generalizing the procedure in Refs.~\cite{Adshead:2013zfa,Creminelli:2011rh}.

		We first rewrite the squeezed-contributing action in terms of the quadratic Hamiltonian density
		\begin{equation}\Ha_2 = a^3 Q \left[ \dot{\zeta}^2 + \frac{c_s^2}{a^2} \left(\pa \zeta\right)^2 \right],\end{equation}
		and the quadratic Lagrangian density 
		\begin{equation}\L_2 = a^3 Q \left[ \dot{\zeta}^2 - \frac{c_s^2}{a^2} \left(\pa \zeta\right)^2 \right],\end{equation}
		such that
		\begin{align*}
		\label{eq:squeezeactionH2L2}
		S_{\text{squeezed}} =& \int \diffcubed{x} \diff{t} \frac{\zeta}{2 Q} \biggl[\left(\Ha_2 + \L_2\right) F_1 \\
							&+  \left(\Ha_2 - \L_2\right) \frac{F_2}{c_s^2} \biggr].
		\numberthis
		\end{align*}

		Next we note that several terms can be grouped into a vanishing boundary term. For a general function of time $F$,
		\begin{align*}
		\label{eq:squeezegrouprelation}
			\frac{1}{F} \frac{\diff{}}{\diff{t}} \left(\frac{F \zeta \Ha_2}{H}\right) =& \ \frac{\dot{\zeta}}{H} \L_2 - \zeta (\Ha_2 + 2 \L_2) - (q+\sigma) \zeta  \L_2 \\
			&+ \left(\frac{\dot{F}}{H F} + \e + \sigma\right) \zeta \Ha_2.
			\numberthis
		\end{align*}

		Ref.~\cite{Adshead:2013zfa} uses a similar relation with $F = 1/c_s^2$ to simplify the action in $k$-inflation. Here we generalize this grouping using
		\begin{equation}
		F = \frac{1}{2+q+\sigma} \left(2 - \frac{F_1}{2Q} + \frac{F_2}{2 c_s^2 Q}\right),
		\end{equation}
		such that the total $\zeta \L_2$ term on the right-hand side of Eq.~\eqref{eq:squeezegrouprelation} corresponds to the $\zeta \L_2$ term in Eq.~\eqref{eq:squeezeactionH2L2}, plus an additional factor of $2 \zeta \L_2$. 

		Making this substitution and using the specific functional forms of $F_1$ and $F_2$, we find a significant cancellation among the $\zeta \Ha_2$ terms which results in the squeezed action taking the form
		\begin{align*}
			S_{\text{squeezed}} =& \int \diff{t} \diffcubed{x} \left[\zeta (\Ha_2 + 2 \L_2) - \frac{F}{H} \dot{\zeta} \L_2 \right. \\&+ \left.\frac{\diff{}}{\diff{t}}\left(\frac{F}{H} \zeta  \Ha_2 \right) \right].
			\numberthis
		\end{align*}
		
		The boundary term here does not contribute to the bispectrum (see Ref.~\cite{Adshead:2013zfa}), and therefore we discard it. The $\dot{\zeta} \L_2$ term does not contribute to the squeezed limit. In order to make the consistency relation more manifest, we undo the grouping by using Eq.~\eqref{eq:squeezegrouprelation} with $F=1$. We also use
		\begin{equation}
		2 G \zeta \L_2 = \frac{\diff{}}{\diff{t}} (G a^3 Q \zeta^2 \dot{\zeta}) - \dot{G} a^3 Q \zeta^2 \dot{\zeta},
		\end{equation}
		which holds for all functions of time $G$, and in particular we use it with $G =  \e + 3\sigma/2 + q/2$.

		After these substitutions and including the terms in Eq.~\eqref{eq:cubiclagrangiandirect} that do not contribute to the squeezed limit, we obtain the cubic action for metric perturbations
		\begin{align*}
		\label{eq:cubicaction}
			S_3 =& \int \diffcubed{x} \diff{t} \biggl[a^3 Q \frac{\diff{}}{\diff{t}} \left(\e + \frac{3}{2}\sigma + \frac{q}{2}\right) \zeta^2 \dot{\zeta} 
		     \\&- \frac{\diff{}}{\diff{t}} \left[ a^3 Q \left(\e + \frac{3}{2}\sigma + \frac{q}{2}\right) \zeta^2 \dot{\zeta}\right] 
		     \\&+ (\sigma + \e) \zeta(\Ha_2 + 2 \L_2) + (1 - F) \frac{\dot{\zeta} \L_2}{H}
			 \\&+ a^3 \frac{F_3}{{H}} \dot\zeta {}^3
			 + a^3 F_4 \dot\zeta  \pa_a\zeta  \pa_a\chi 
			 \\&+ a^3 F_5 \pa^2\zeta  \left(\pa \chi \right)^2 
			 + \frac{F_6}{{H^3} a} \dot\zeta  \pa^2\zeta  \pa^2\zeta
			 \\&+ \frac{F_7}{{H^4} a^3} \pa^2\zeta \left(\pa_a\pa_b\zeta \right)^2
			    + \frac{F_8}{{H^4} a^3} \pa^2\zeta  \pa^2\zeta  \pa^2\zeta
			    \\&+ \frac{F_9}{{H^3} a} \pa^2\zeta (\pa_a\pa_b\zeta)(\pa_a\pa_b\chi) \biggr].
			 \numberthis
		\end{align*}
		
		Refs.~\cite{Adshead:2013zfa,Creminelli:2011rh} show explicitly in the context of more restricted inflationary models that the boundary term yields the slow-roll squeezed-limit consistency relation, while the first term on the first line contributes to the squeezed-limit at higher order in slow-roll, as does the first term on the third line (which can be seen by re-application of Eq.~\eqref{eq:squeezegrouprelation}). No other term contributes to the squeezed limit at lowest order in slow-roll, and therefore we can immediately see from Eq.~\eqref{eq:cubicaction} that the squeezed-limit consistency relation holds in the unified EFT of inflation during slow-roll. In \S\ref{sec:GSR}, we will show that the consistency relation holds even beyond slow-roll.

		While the cubic action \eqref{eq:cubicaction} ensures the consistency relation holds in slow-roll, no assumption of slow-roll has been made in its derivation. 

	\subsection{Horndeski and GLPV subclasses}
		\label{subsec:HorndeskiGLPV}

		Though we write the EFT directly in terms of  the metric, the EFT can also be viewed as a four-dimensional scalar-tensor theory by transforming out of unitary gauge using the Stuckelburg trick \cite{Gao:2014soa,Cheung:2007st}. In this way, the EFT of inflation presented in \S\ref{subsec:EFTLagrangian} encompasses a large space of fully covariant models. In this section, we study the structure of the cubic action \eqref{eq:cubicaction} derived in \S\ref{subsec:CubicActionConsistency} in the Horndeski and GLPV model classes.

		The Horndeski and GLPV classes are constructed to avoid the Ostrogradsky instability \cite{Woodard:2015zca,Solomon:2017nlh}. The Horndeski class \cite{Horndeski:1974wa} is the most general four-dimensional scalar-tensor theory with second-order equations of motion for the scalar field $\phi$. The Horndeski class can be broadened to include models which have higher than second-order equations of motion yet due to a degeneracy condition do not propagate an Ostrogradsky mode. This is  the beyond-Horndeski GLPV class \cite{Gleyzes:2014qga}, of which the Horndeski class is a subset. The GLPV class is an example of a DHOST theory \cite{BenAchour:2016fzp}. While the GLPV model can be represented with an action of the form \eqref{eq:EFTAction}, writing the other DHOST theories in our EFT would require generalizing Eq.~\eqref{eq:EFTAction} to include time derivatives of the lapse function \cite{Langlois:2017mxy}.

		The cubic action \eqref{eq:cubicaction} and the resultant bispectrum takes on a restricted form in the Horndeski and GLPV classes. This restriction follows from the ADM representation of the  action for Horndeski and GLPV models \cite{Kase:2014cwa},
		\begin{align*} 
			L =& \ A_2 + A_3 K 
			+ A_4 (K^2 - K^i_{\hphantom{i}j} K^j_{\hphantom{i}i})  + B_4  R
			\\
			& + A_5 ( K^3 - 3K K^i_{\hphantom{i}j} K^j_{\hphantom{i}i} + 2 K^i_{\hphantom{i}j} K^j_{\hphantom{i}k} K^k_{\hphantom{i}i} ) \\
			& + B_5 ( K^i_{\hphantom{i}j}R^j_{\hphantom{i}i}-\tfrac{1}{2}KR ).
			\numberthis
		\end{align*}
		Here  $A_n(X,\phi)$ and $B_n(X,\phi)$ are functions of the kinetic term $X= \nabla^\mu\phi \nabla_\mu\phi$ and field $\phi$.  In the unitary gauge of ADM,  $\phi\rightarrow\phi(t)$  and thus $X = -\dot{\phi}^2/ N^2$, so these quantities may also be considered
		as functions of $N$ and $t$. In the GLPV class, these functions are completely general, while in the Horndeski class they satisfy
		\begin{align*}
		A_4 &= 2 X B_{4,X} - B_4,\\
		A_5 &= - \frac{1}{3} X B_{5,X}.
		\numberthis
		\end{align*}
		
		We then take the appropriate partial derivatives in Eq.~\eqref{eq:cubicADMLagrangian} to get the various $\C$ variables in the Horndeski and GLPV theories. We find
		\begingroup
		\allowdisplaybreaks
		\begin{align*}
		 \C_{N} =&
		 - 2 X (A_{2,X}\!
		 + 3 A_{3,X} H
		 + 6 A_{4,X} H^2\!
		 + 6 A_{5,X} H^3 ), \\
		\C_{R}
		 =& \  B_{4}
		 - \frac{B_{5} H}{2},\\
		\C_{NN} =& \ 6 X \left(A_{2,X}
		 + 3 A_{3,X} H 
		 + 6 A_{4,X} H^2
		 + 6 A_{5,X} H^3\right) \\
		 &+ 4 X^2 \left(A_{2,XX}
		 + 3 A_{3,XX} H
		 + 6 A_{4,XX} H^2 \right. \\
		 &+ \left. 6 A_{5,XX} H^3\right), \\
		\C_{K} =& \  A_{3}
		 + 4 A_{4} H
		 + 6 A_{5} H^2, \\
		\C_{KK}
		 =& \ 2 (A_{4}
		 + 3 A_{5} H), \\
		\C_{NK}
		 =&  
		 - 2 \left(A_{3,X}
		 + 4 A_{4,X} H
		 + 6 A_{5,X} H^2\right) X ,\\
		\C_{NR}
		 =& 
		 - 2 \left(B_{4,X}
		 - \frac{B_{5,X} H}{2}\right) X ,\\
		\C_{KR}
		 =& - \frac{B_{5}}{2},\\
		\C_{NNN}
		 =&  
		 - 24 X \left(A_{2,X}
		 + 3 A_{3,X} H
		 + 6 A_{4,X} H^2 \right. \\
		 &+ \left. 6 A_{5,X} H^3\right)\\
		 &- 36 X^2 \left(A_{2,XX}
		 + 3 A_{3,XX} H
		 + 6 A_{4,XX} H^2 \right. \\
		 &+ \left. 6 A_{5,XX} H^3\right)\\
		 &- 8 X^3 \left(A_{2,XXX}
		 + 3 A_{3,XXX} H
		 + 6 A_{4,XXX} H^2 \right. \\
		 &+ \left. 6 A_{5,XXX} H^3\right), \\
		\C_{NNK}
		 =&  \ 6 X \left(A_{3,X}
		 + 4 A_{4,X} H
		 + 6 A_{5,X} H^2\right) \\
		 &+ 4 X^2 \left(A_{3,XX}
		 + 4 A_{4,XX} H
		 + 6 A_{5,XX} H^2\right),\\
		\C_{NNR}
		 =&  \ 6 X \left(B_{4,X}
		 - \frac{B_{5,X} H}{2}\right)  \\
		 &+ 4 X^2 \left(B_{4,XX}
		 - \frac{B_{5,XX} H}{2}\right) ,\\
		\C_{NKK}
		 =&
		 - 4 X (A_{4,X}
		 + 3 A_{5,X} H) ,\\
		\C_{N\bar{K}\bar{K}}
		 =& \ 
		  4 X (
		   A_{4,X}
		  +3 A_{5,X} H) ,\\
		 \C_{NKR} =& \ X B_{5,X}, \\
		 \C_{N\bar{K}\bar{R}} =&  - 2 X B_{5,X}, \\
		 \C_{KKK} =& \ 6 A_{5},\\
		 \C_{\bar{K}\bar{K}K} =& - 6 A_{5}, \\
		 \C_{\bar{K}\bar{K}\bar{K}} =&  \ 12 A_{5},
	  	\numberthis
		\end{align*}
		\endgroup
		in which ${}_{,X}\equiv d/dX$ and all other coefficients are either zero or determined by Eq.~\eqref{eq:QuadraticInheritance}. 
		
		Using these coefficients, one can show that in the Horndeski and GLPV cases $F_6$, $F_7$, $F_8$, and $F_9$ are identically zero using the expressions in Appendix~\ref{app:Fs}. 
		Thus the cubic action reduces to
		\begin{align*}
		\label{eq:cubicactionHorndeski}
			S^{\text{GLPV}}_3 =& \int \diffcubed{x} \diff{t} \biggl[a^3 Q \frac{\diff{}}{\diff{t}} \left(\e + \frac{3}{2}\sigma + \frac{q}{2}\right) \zeta^2 \dot{\zeta} 
		     \\&- \frac{\diff{}}{\diff{t}} \left[ a^3 Q \left(\e + \frac{3}{2}\sigma + \frac{q}{2}\right) \zeta^2 \dot{\zeta}\right] 
		     \\&+ (\sigma + \e) \zeta(\Ha_2 + 2 \L_2) + (1 - F) \frac{\dot{\zeta} \L_2}{H}
			 \\&+ a^3 \frac{F_3}{{H}} \dot\zeta {}^3
			 + a^3 F_4 \dot\zeta  \pa_a\zeta  \pa_a\chi 
			 \\&+ a^3 F_5 \pa^2\zeta  \left(\pa \chi \right)^2 \biggr].
			 \numberthis
		\end{align*}

		It can also be shown that the $F_4$ and $F_5$ operators are suppressed by an additional factor of slow-roll parameters relative to the other operators. This result was shown for the Horndeski class in Ref.~\cite{DeFelice:2013ar}, and holds also for the GLPV class.

		This is the same form of the action as shown in Refs.~\cite{RenauxPetel:2011sb,DeFelice:2011uc,Gao:2011qe}, after undoing our grouping of the $F_1$ and $F_2$ terms. Our novel squeezed-action grouping of $F_1$ and $F_2$ also confirms the result of Ref.~\cite{DeFelice:2013ar} that the squeezed-limit consistency relation holds in slow-roll in Horndeski models
		and corroborates the result in Ref.~\cite{Fasiello:2014aqa} that GLPV leads to no new scalar bispectrum shapes relative to Horndeski. By writing it in this form we show that the squeezed-limit consistency relation holds in GLPV models in slow-roll.

\section{Generalized Slow-Roll Bispectrum}
	\label{sec:GSR}
	We present in \S\ref{subsec:InInGSRFormalism} the in-in and generalized slow-roll formalisms, which we use to construct a complete integral formulation of the bispectrum beyond slow-roll resulting from the cubic EFT action derived in \S\ref{sec:EFT}. In \S\ref{subsec:squeeze}, we study the squeezed-limit of the bispectrum and show that the consistency relation holds beyond slow-roll. We relegate the explicit forms for the components to Appendix~\ref{app:EFTIntegrals}.

	\subsection{In-In and GSR formalisms}
		\label{subsec:InInGSRFormalism}

		The tree-level three-point correlation function in the in-in formalism is given by \cite{Adshead:2013zfa,Maldacena:2002vr,Weinberg:2005vy,Adshead:2009cb}
		\begin{align*}
		\label{eq:inin}
		&\langle \hat{\zeta}_{\bf{k_1}}(t_*) \hat{\zeta}_{\bf{k_2}}(t_*) \hat{\zeta}_{\bf{k_3}}(t_*) \rangle = \numberthis \\
		&2  \re{-i \int_{-\infty(1+i\epsilon)}^{t_*} \diff{t} \langle \hat{\zeta}^I_{\bf{k_1}}(t_*) \hat{\zeta}^I_{\bf{k_2}}(t_*) \hat{\zeta}^I_{\bf{k_3}}(t_*) H_{I}(t) \rangle},
		\end{align*} 
		with $H_I \simeq - \int \diffcubed{x} \L_3$ at cubic order \cite{Adshead:2008gk}. 

		The field operators $\hat{\zeta}^I$ are in the interaction picture, which means their corresponding modefunctions satisfy the free Hamiltonian's equation of motion \eqref{eq:eom}. $\hat{\zeta}_{\bf k}^I$ is the Fourier transform of the operator. 
		We define the corresponding modefunctions $\zeta_k(t)$ as
		\begin{equation}
		\hat{\zeta}^I_{\bf{k}}(t) = \zeta_k(t) \hat a({\bf k}) + \zeta_k^* \hat a^\dag(-{\bf k}) ,
		\end{equation}
		where the creation and annihilation operators satisfy
		\begin{equation}
		[\hat a({\bf k}), \hat a^\dag({\bf k}')] = (2\pi)^3 \delta({\bf k}-{\bf k}')
		\end{equation}
		as usual.    Using these relations  the power spectrum can be evaluated from the modefunctions at a time $t_*$
		taken to be after all the relevant modes have left the horizon 
			\begin{align}
		\langle \hat{\zeta}^I_{\textbf{k}}(t_*)\hat{\zeta}^I_{\textbf{k}'}(t_*)\rangle & = 
		 (2\pi)^3\delta^{3}(\textbf{k}+\textbf{k}') |\zeta_k(t_*)|^2  
		\nonumber \\ &\equiv  (2\pi)^3\delta^{3}(\textbf{k}+\textbf{k}') P_\zeta (k). 
		 \end{align}

		Translational and rotational invariance requires that the three-point correlators be encapsulated in the bispectrum $B_\zeta$ as
		\begin{equation}
		\langle \hat{\zeta}_{\bf{k_1}} \hat{\zeta}_{\bf{k_2}} \hat{\zeta}_{\bf{k_3}}\rangle
		= (2\pi)^3 \delta^{3}({\bf{k_1}}+{\bf{k_2}}+{\bf{k_3}}) B_{\zeta}(k_1,k_2,k_3), 
		\end{equation}
		in which we have suppressed the evaluation at $t_*$. The dimensionless parameter conventionally constrained by experiment is
		\begin{equation}
		f_{\textrm{NL}} (k_1, k_2, k_3) \equiv \frac{5}{6} \frac{B_\zeta(k_1, k_2, k_3)}{P_\zeta(k_1) P_\zeta(k_2) + \text{perm.}} \ \ .
		\end{equation}
		Here and throughout `$+\text{perm.}$' denotes the two additional cyclic permutations of indices.

		In order to evaluate the in-in integral \eqref{eq:inin} and compute $B_{\zeta}(k_1,k_2,k_3)$, we need to solve the equation of motion \eqref{eq:eom} for the interaction picture modefunctions $\zeta_k(t)$.
		However, beyond the slow-roll approximation, there is no general analytic solution to the equation of motion. The generalized slow-roll approach is to solve the equation of motion iteratively \cite{Adshead:2013zfa,Stewart:2001cd,Choe:2004zg,Dvorkin:2009ne,Hu:2011vr,Kadota:2005hv}.
		It is convenient to express the modefunction in dimensionless form as
		\begin{equation}
		y \equiv \sqrt{\frac{k^3}{2\pi^2} } \frac{f}{x} \zeta_k,
		\end{equation}
		where
		\begin{equation}
		f \equiv 2 \pi a s \sqrt{2 Q c_s},
		\end{equation}
		$x= k s$ and the sound horizon 
		\begin{equation}
		s \equiv \int_a^{a_\text{end}} \frac{ \diff{\tilde{a}}}{\tilde a} \frac{c_s}{\tilde{a} H},
		\end{equation}
		with $a_{\rm end}$ denoting the end of inflation.
		
		The formal solution to Eq.~\eqref{eq:eom} is 
		\begin{equation}
		\label{eq:modefunctionformal}
		y(x) = y_0 (x) - \int_x^\infty \frac{\diff{\tilde x}}{\tilde x^2} g(\ln \tilde{s}) y(\tilde x) \im{y_0^*(\tilde x) y_0(x)},
		\end{equation}
		in which  $\tilde x \equiv k \tilde{s}$, $g(\ln s) \equiv (f'' - 3f')/f$	and $' \equiv d/d \ln s$.
		The zeroth order solution with Bunch-Davies initial conditions is 
		\begin{equation}
		y_0 (x) = \left(1+\frac{i}{x}\right) e^{i x}.
		\end{equation}
		The first-order solution is obtained by plugging in the zeroth-order solution into the right-hand side of Eq.~\eqref{eq:modefunctionformal}.

		Every order in the GSR hierarchy of solutions is suppressed relative to the previous order by the $g$ factor, whose time integral is assumed to be small but whose value can evolve and become transiently large unlike in the slow-roll approximation -- we call such a case ``slow-roll suppressed". When operators in the cubic action are
		also slow-roll suppressed, as is the case for the $\zeta^2 \dot{\zeta}$ and $\zeta(\Ha_2 + 2 \L_2)$ terms, it suffices to use the zeroth-order solution for the modefunctions in computing the bispectrum to first order in slow-roll parameters. Operators with general EFT coefficients, however, are not necessarily slow-roll suppressed and therefore the first-order modefunction solution must be used in order to maintain a consistent first-order solution.  

		In the GSR formalism, the power spectrum to first order in slow-roll parameters is \cite{Miranda:2012rm,Motohashi:2017gqb}
		\begin{equation}
		\label{eq:GSRpower}
		\ln \Delta_\zeta^{2} = G(\ln s_{*}) + \int_{s_{*}}^\infty {\frac{\diff s}{s}} W(ks) G'(\ln s),
			\end{equation}
		with the power spectrum window function
		\begin{equation}
		W(u) = {\frac{3 \sin(2 u)}{2 u^3}} - {\frac{3 \cos (2 u)}{u^2}} - {\frac{3 \sin(2 u)}{2 u}},
		\end{equation}
		and the power spectrum source
		\begin{equation}
		G = - 2 \ln f    + \frac{2}{3} (\ln f )'.
		\end{equation}

		The first-order bispectrum result follows the same schematic form as the first-order power spectrum result: a windowed integral over a source. Each operator $i$ in the cubic action contributes a set of sources and windows to the bispectrum which are indexed by $j$ according to their asymptotic scalings at $x\ll1$ and $x\gg 1$. Thus we denote these sources and windows as $S_{ij}$, $W_{ij}$. 

		At zeroth order in GSR modefunctions, the bispectrum integrals depend only on the triangle perimeter $K \equiv k_1+k_2+k_3$, and all shape dependence is held outside the integrand by corresponding $k$-weights  $T_{ij}$. The integrals take the form
		\begin{align*}
		\label{eq:GSRIntegration}
		 I_{ij}(K) = S&_{ij}(\ln s_*) W_{ij}(K s_*) \\&+ \!\int_{s_*}^\infty \!\frac{\diff s}{s}  S_{ij}'(\ln s) W_{ij}(K s).\numberthis
		\end{align*}
		
		At first order in the GSR modefunctions, each operator yields a shape-dependent boundary contribution resulting from the removal of certain nested integrals using integration by parts \cite{Adshead:2013zfa}. These contributions are of the form $\left[T_{i B} I_{i B} (2 k_3) + \text{perm.}\right]$.
		Together, the perimeter-dependent and shape-dependent integrals enable computation of the complete bispectrum of the effective field theory of inflation to first order in slow-roll parameters,
		\begin{align*}
		\label{eq:GSRBi}
		B_{\zeta}&(k_1,k_2,k_3) =  \frac{(2 \pi)^4}{4} \frac{\Delta_\zeta(k_1) \Delta_\zeta(k_2) \Delta_\zeta(k_3)}{k_1^2 k_2^2 k_3^2} \numberthis\\ &\times \Big\{  \sum_{ij} T_{ij} I_{ij}(K) +
		 \sum_{i=2}^9 \left[T_{iB} I_{iB}(2 k_3) +{\rm perm.}\right] \Big\}.
		\end{align*}
		
		We provide the sources, windows, and $k$-weights that each operator in the cubic action contributes to this expression in Appendix~\ref{app:EFTIntegrals}. In Table~\ref{tab:OperatorSummary}, we give summary information for each operator. The following section focuses on establishing the squeezed-limit consistency relation beyond slow-roll from these results.

		\begin{table}[tb]
		\begin{center}
		\begin{tabular}{ l  c c  c c }
		$i$  &  Operator &  Source & Squeezed & GLPV \\
		 \hline
		 &&&&\\[-10pt]
		0 \ {} & $\zeta^2 \dot \zeta$ & $ \dfrac{2 \e + 3 \sigma + q }{2 f}$ & yes &  Supp. \\[10pt]
		1 & $\zeta({\Ha}_2+2 {\L}_2)$  & $\dfrac{\sigma+\e}{f}$ & yes &  Supp.  \\[10pt]
		2 & $\dot \zeta {\cal L}_2$  & $ \dfrac{c_s}{a H s} \dfrac{F-1}{f}$ & no & Free \\[10pt]
		3 & $\dot \zeta^3$ & $- \dfrac{1}{Q} \dfrac{c_s}{a {H} s} \dfrac{F_3}{f}$ & no &  Free \\[10pt]
		4 & $ \dot\zeta (\partial\zeta)  \pa \chi  $ & $-\dfrac{1}{2Q}\dfrac{F_4}{f}$ & no & Supp. \\[10pt]
		5 & $ \pa^2 \zeta (\pa \chi)^2  $ & $\dfrac{1}{Q}\dfrac{F_5}{f}$ & no & Supp. \\[10pt]
		6 & $\dot{\zeta} \pa^2 \zeta \pa^2 \zeta$ & $ { \dfrac{1}{Q c_s^4}  \left( \dfrac{c_s}{a Hs} \right)^3}
		\dfrac{F_6}{f}$ & no & -- \\[10pt]
		7 & $(\pa_a \pa_b \zeta)^2 \pa^2 \zeta$ & 
		$
		{ \dfrac{1}{Q c_s^6}  \left( \dfrac{c_s}{a Hs} \right)^4}
		\dfrac{F_7}{f}$ & no & -- \\[10pt]
		8 & $(\pa^2 \zeta) (\pa^2 \zeta) (\pa^2 \zeta)$ & ${ \dfrac{1}{Q c_s^6}  \left( \dfrac{c_s}{a Hs} \right)^4}
\dfrac{F_8}{f}$ & no & -- \\[10pt]
		9 & $(\pa^2 \zeta) (\pa_a \pa_b \zeta) (\pa_a \pa_b \chi)$ & ${ \dfrac{1}{Q c_s^4}  \left( \dfrac{c_s}{a Hs} \right)^3}
\dfrac{F_9}{f}$ & no & -- \\[10pt]
		\end{tabular}
		\end{center}
		\caption{GSR bispectrum operators, sources, whether they contribute to the squeezed-limit, and their status in the GLPV class and its subset the Horndeski class. ``Supp.'' denotes an operator which is slow-roll suppressed, while ``Free'' an operator which is not. The $i=6$, $i=7$, $i=8$, and $i=9$ operators are identically zero in the GLPV and Horndeski classes.}
		\label{tab:OperatorSummary}
		\end{table}

	\subsection{Consistency relation}
		\label{subsec:squeeze}

		In \S\ref{subsec:CubicActionConsistency}, we argued from the cubic action that the squeezed-limit consistency relation \eqref{eq:ConsistencyRelation} holds during slow-roll inflation. Now that we have the complete integral forms of the bispectrum to first order in slow-roll parameters, we can examine the squeezed-limit consistency relation in more detail, in particular focusing on its form beyond slow-roll.

		We first confirm our expectation from \S\ref{subsec:CubicActionConsistency} that only the
		 $i=0$ and $i=1$ operators  contribute in the squeezed-limit. In Appendix~\ref{app:EFTIntegrals}, we show that in the squeezed-limit $x_L / x_S \gg 1$, $x_S \ll 1$, we have that
		\begin{equation}
		\sum_{i=2}^{9} \Bigl[ \sum_j T_{ij} I_{ij} + \left[T_{iB} I_{iB}(2 k_3) + \text{perm.}\right] \Bigr] = 0,
		\end{equation}
		and therefore the operators $i=2$ to $9$ have no net squeezed contribution.

		As for the $i=0$ and $i=1$ operators, only  $I_{01}$, $I_{02}$, $I_{11}$, and $I_{12}$ contribute to squeezed triangles as $k_L/k_S$.   We can then generalize a calculation from Ref.~\cite{Adshead:2013zfa} to show that the consistency relation holds even beyond slow-roll. The GSR expression for the squeezed bispectrum is
		\begin{align*}
		\frac{12}{5} f^{\text{squeezed}}_{\text{NL}} = \lim_{k_S \rightarrow 0} &\frac{1}{\Delta (k_S)} \Bigl[-2 I_{01}(2 k_L)  + 4 I_{02}(2 k_L) \\ &+ 2 I_{11}(2 k_L) - 2 I_{12}(2 k_L) \Bigr]. \numberthis
		\end{align*}

		To leading order we can substitute $1/\Delta(k_S) \rightarrow f_*$, where $s_*$ is an epoch during slow-roll, resulting in
		\begin{align*}
		\label{eq:GSRconsistency}
		\frac{12}{5} f_{\text{NL}}
		\approx&
		-2 \frac{f'}{f}\Big|_{s_*} 
		+  f_* \int_{s_*}^\infty \frac{d s}{s}   \bigg[ \left( \frac{\e}{f} \right)' W_{\epsilon}( k_L s) \numberthis \\ &  \qquad 
		+
		\left( \frac{\sigma}{f} \right)' W_{\sigma}( k_L s) 
		+
		\left( \frac{q}{f} \right)' W_{q}( k_L s) \bigg], 
		\end{align*}
		where
		\begin{align*}
		 W_{\epsilon}(x) =& \ \frac{1}{x}  \sin(2 x),\\
		 W_{\sigma}(x) =& \ \frac{2}{x}  \sin(2 x) - \cos (2 x),\\
		 W_{q}(x) =& \ \frac{1}{x}  \sin(2 x) - \cos (2 x),
		 \numberthis
		 \end{align*}
		and we have evaluated the boundary term during slow-roll as
		\begin{equation}
		(2 \e + 3 \sigma + q)|_{s_*} \simeq -2 \frac{f'}{f}\Big|_{s_*}.
		\end{equation}

		We need to compare this GSR expression for the squeezed bispectrum to the GSR expression for the tilt of the power spectrum \cite{Adshead:2012xz},
		\begin{align*}
		\label{eq:slope}
		\frac{ d \ln \Delta_\zeta^2}{d \ln k}\Big|_{k_L}  &= \int_{s_*}^\infty \frac{ds}{s} W'(k_L s) G'(\ln s)
		\numberthis\\
		&=
		2 \frac{f'}{f}\Big|_{s_*} + \int_{s_*}^\infty \frac{ds}{s} \left( \frac{f'}{f} \right)' W_{n}(k_L s),
		\end{align*}
		where
		\begin{equation}
		W_{n}(x) =-2 \cos(2 x) + \frac{2}{x} \sin(2x).
		\end{equation}

		We see immediately from comparing the boundary terms in Eqs.~\eqref{eq:GSRconsistency} and \eqref{eq:slope} that the squeezed limit consistency relation holds in slow-roll. The integral contributions become significant during slow-roll violations. For a sharp feature at $k_L s \gg 1$, the parameters with the highest numbers of derivatives dominate and
		\begin{equation}
		\label{eq:sharpstepappx}
		\left( \frac{f'}{f} \right)'  \approx \frac{f''}{f} \approx - \frac{f_*}{2} \left( \frac{\sigma + q}{f} \right)',
		\end{equation}
		which, when combined with the windows in the desired limit, establishes consistency beyond slow-roll between Eqs.~\eqref{eq:GSRconsistency} and \eqref{eq:slope}.

		We have made two assumptions in deriving the consistency relation beyond slow-roll. First, we have assumed that the net change in the power spectrum between two different scales
		is slow-roll suppressed and thus that we can send $1/ \Delta (k_S)$ to $f_*$.  Implicitly this requires that 
		any slow-roll violation is highly transient so that the integrated effect of transient violations remains small.   Therefore
		second, we assume that the sources of slow-roll violation are sharp in their temporal structure using Eq.~\eqref{eq:sharpstepappx}. The inflationary model we consider in the following section can violate these approximations by allowing large changes in the power spectrum outside the well observed regime. Nonetheless, we expect that the consistency relation when computed exactly holds in general as long as $\zeta$ freezes out after
		horizon crossing.		
		
\section{Transient G-Inflation}
	\label{sec:GInflation}
	In this section, we illustrate the
	calculation of  the scalar bispectrum in our general formalism for the unified EFT of inflation with a specific  model with cubic Galileon interactions
	  in which slow-roll is transiently violated.  We briefly review this transient G-inflation model in \S\ref{subsec:GInfModel} and present its bispectrum in \S\ref{subsec:GInfResults}.

	\subsection{Model}
		\label{subsec:GInfModel}
		The transient G-inflation model is presented in detail along with its scalar and tensor power spectra in Ref.~\cite{Ramirez:2018dxe}. We briefly review it here.

		We assume that the Lagrangian density takes the form 
		\begin{equation}
		\L = -X/2 - V(\phi) + f_3(\phi) \frac{X}{2} \Box \phi + \frac{R}{2},
		\end{equation}
 		with the chaotic inflation potential $V(\phi) = m^2 \phi^2/2$. In Ref.~\cite{Ohashi:2012wf}, this model is considered with a constant $f_3 = -M^{-3}$. The constant $f_3$ model suffers from two problems: for the measured value of the scalar tilt $n_s$, it predicts too large a tensor-to-scalar ratio $r$; and for some values of $m$ and $M$ the inflaton has a gradient instability $c_s^2 < 0$ during reheating whose resolution would lie beyond the scope of the perturbative EFT.

		Transient G-inflation shuts off the G-inflation term before the end of inflation by using a $\tanh$ steplike feature in $f_3$
		\begin{equation}
		f_3 (\phi) = -M^{-3} \left[ 1 + \tanh\left(\frac{\phi-\phi_r}{d}\right)\right].
		\end{equation}
		
		Prior to the step, the inflaton is in a G-inflation regime, while after the step, the inflaton follows the slow-roll attractor solution of chaotic inflation. Because the $f_3 X \Box \phi$ term in the Lagrangian becomes negligible after the step, the gradient instability at the end of inflation is avoided. By having the transition start just as the CMB scale exits the horizon, the tilt $n_s$ is decoupled from the tensor-to-scalar ratio $r$ and therefore the model can be consistent with observations.

		We consider two parameter sets for the transient G-inflation model, a `large-step' model and a `small-step' model.
		The large-step model is the fiducial model of Ref.~\cite{Ramirez:2018dxe}. The inflaton mass scale $m = 2.58 \times 10^{-6}$ is chosen to satisfy the {\it Planck} 2015 TT+lowP power spectrum amplitude. The Galileon mass scale $M=1.303 \times 10^{-4}$ suppresses the tensor amplitude relative to the scalar amplitude when the CMB mode $k_{\text{CMB}} = 0.05$\,Mpc$^{-1}$ exits the horizon $55$ $e$-folds before the end of inflation. The remaining parameters $\phi_r=13.87$ and $d=0.086$ control the step and are chosen such that the tilt and running satisfy observational constraints.

		The small-step model is chosen by the same procedure, save for the parameter $M$ which is selected for a larger tensor amplitude. The other parameters are adjusted to keep the tilt and amplitude of the power spectrum fixed. The resultant parameter set is $\{m, M, \phi_r, d\} = \{6.50 \times 10^{-6},\ 48.25 \times 10^{-4},\ 14.67,\ 0.021\}$. In this model, the power spectrum evolution before and after the step is much smaller as inflation is never in a fully G-inflation dominated phase, and thus we are closer to the regime of validity of the argument in \S\ref{subsec:squeeze}.

		\begin{figure}[t]
		\psfig{file=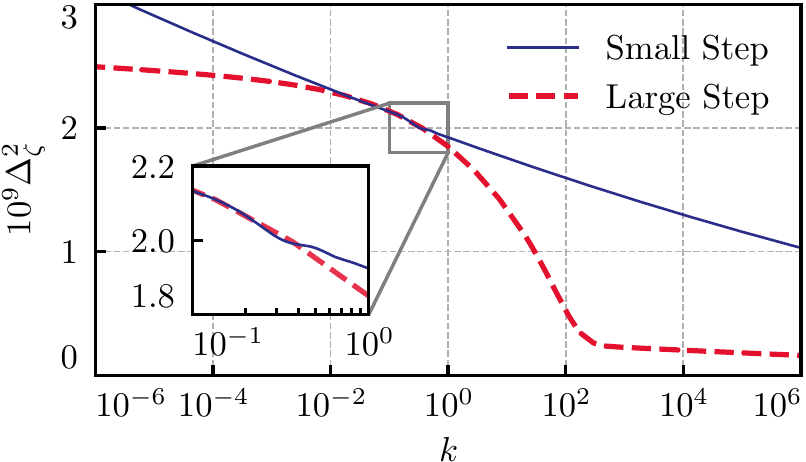}
		\caption{The GSR power spectra for the transient G-inflation models we consider.  In the small step model, the transition has a small amplitude but rapid variation 
		whereas in the large step model, it has a large amplitude and slow variation.}
		\label{fig:powerSpectra}
		\end{figure}

		In both models, slow-roll is transiently violated as the inflaton traverses the step, and thus the GSR formalism should be used in place of the traditional slow-roll approach for power spectrum and bispectrum observables. We show the GSR power spectra for these models in Fig.~\ref{fig:powerSpectra}.  In the small-step model, the deviations from scale invariance 
		are small in amplitude but rapidly varying in $k$ (see inset).   In the large-step model, they
		are large in amplitude but smoother in scale.    We shall see next that these properties 
		also apply to the bispectrum.

	\subsection{GSR bispectrum for transient G-Inflation}
		\label{subsec:GInfResults}

		We now compute the bispectrum for the transient G-inflation models of \S\ref{subsec:GInfModel} using the GSR formulas from \S\ref{sec:GSR}.
		
		We begin by computing the squeezed bispectrum, where the consistency relation allows us to check our computations by comparing the bispectrum result in the squeezed limit to the slope of the GSR power spectrum using Eq.~\eqref{eq:ConsistencyRelation}. We choose to fix the ratio $k_S / k_L = 10^{-2}$. From the analytic analysis in \S\ref{subsec:squeeze}, we know that the only operators which contribute to the squeezed limit are the $i=0$ and $i=1$ operators, and their sources are manifestly related to the local slope of the power spectrum. Thus we expect these operators to enforce the consistency relation. 

		\begin{figure}[t]
		\psfig{file=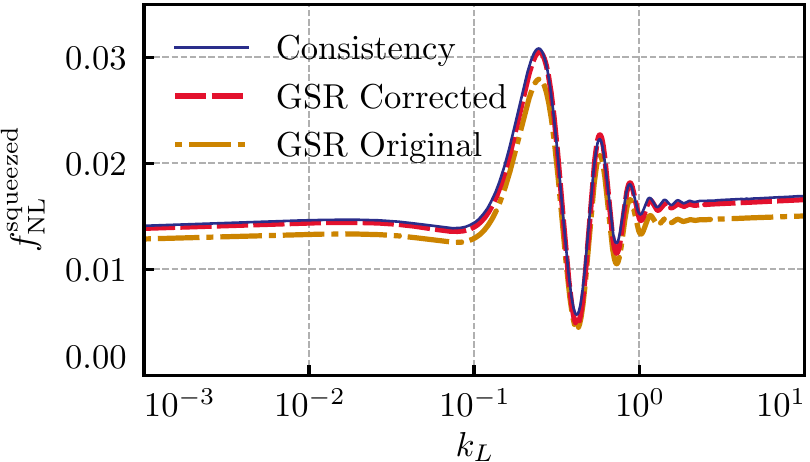}
		\caption{Squeezed bispectrum for small-step transient G-inflation. We see excellent agreement between the GSR bispectrum and the consistency relation curve, though with a slight amplitude error. By applying a simple correction to account for modefunction evolution outside the horizon, we can eliminate this error completely.}
		\label{fig:shortSqueezed}
		\end{figure}

		The accuracy of the GSR approximation in the squeezed-limit for the small-step case is shown in Fig~\ref{fig:shortSqueezed}. The GSR bispectrum result closely tracks the consistency relation result before, during, and after the step in the power spectrum. Slow-roll violations during the transition appear as sharp features in the sources which, when integrated against the windows, induce oscillatory features in the squeezed bispectrum and in the tilt of the power spectrum. 
		
		While the  GSR bispectrum calculation and the power spectrum based consistency relation expectation agree on the period and phase of these features, there is a small amplitude difference between the curves before, during, and after the transition. This error occurs because the
		bispectrum and power spectrum are calculated to first order in slow-roll suppressed
		quantities.  In particular the
		 consistency relation check of \S\ref{subsec:squeeze} ignores corrections due to the
		 evolution in $f$ which would be picked up in the next order of the GSR iteration. 
		  Since there is some slow-roll suppressed evolution in $f$ between the epochs when
		   $k_S$ and $k_L$ freeze out, or equivalently in the power spectra at the two scales,
		   a correspondingly small error is induced in the bispectrum. 
		   
		   In this case, where the 
		   change in the  power spectrum  between $k_S$ and $k_L$ is insignificant, this error is minor. Nonetheless, in the upcoming large-step example the power spectrum will significantly evolve across freeze-out epochs and this error will become large. In Refs.~\cite{Adshead:2013zfa,Adshead:2012xz}, it is shown that next-order terms in the GSR hierarchy provide a correction factor
		\begin{equation}
		\label{eq:gsr_squeeze_corr}
		R_0 = 1 + \frac{n_s-1}{2} \ln \left(\frac{k_S}{k_L}\right),
		\end{equation}
		assuming that the squeezed bispectrum integrals receive most of their contributions
		at horizon crossing for $k_L$.  

		This correction multiplies the zeroth-order bispectrum contributions from the $i=0$ and $i=1$ terms and corrects for the leading-order integrated evolution of $f$. Since in the following example the power spectrum evolution will be large, we generalize this correction to the nonleading integrated evolution of $f$ by choosing.
		\begin{align}\label{eq:gsr_squeeze_corr_resum}
		R = \Delta(k_S)/\Delta(k_L).
		\end{align}
		
		We show in Fig.~\ref{fig:shortSqueezed} that this correction eliminates the small amplitude error, improving the consistency between the squeezed bispectrum and the derivative of the
		 power spectrum. This correction does not impact triangle shapes where all three modes are comparable in scale. For a formulation of GSR which avoids this type of error by maintaining order-by-order modefunction freeze-out, see Ref.~\cite{Miranda:2015cea}.

		\begin{figure}[t]
		\psfig{file=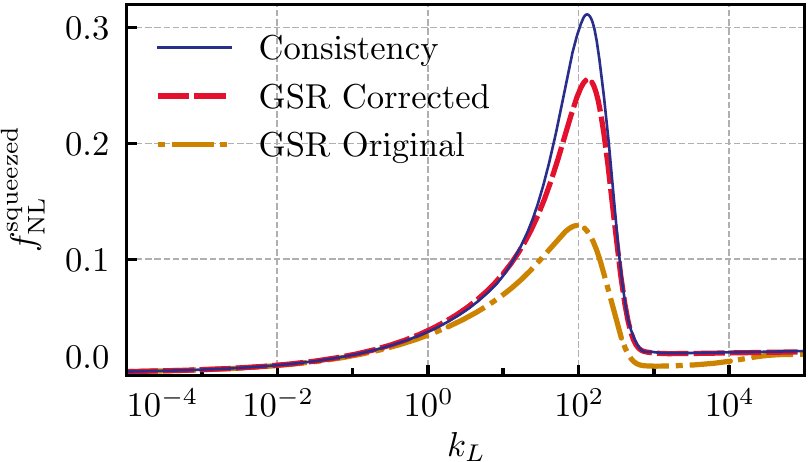}
		\caption{Squeezed bispectrum for large-step transient G-inflation. Despite the large evolution of the power spectrum in this model, the corrected GSR bispectrum tracks closely the consistency relation. The discrepancy between the corrected bispectrum result and the consistency relation at the peak indicates that the next-order term in the GSR hierarchy becomes important there.}
		\label{fig:wideSqueezed}
		\end{figure}

		We show in Fig.~\ref{fig:wideSqueezed} the squeezed bispectrum for the large-step model. In the large-step model, the $i=0$ and $i=1$ sources are much wider than in the small-step case and thus the bispectrum appears as a single peak rather than an oscillatory function. In addition, for this choice of model parameters the power spectrum evolution is large and thus the GSR squeezed bispectrum makes a significant error across the step. Nonetheless, correcting the bispectrum for the integrated evolution $f$ between $k_S$ and $k_L$ with Eq.~\eqref{eq:gsr_squeeze_corr_resum} succeeds in explaining most of this discrepancy.
		
		The residual errors in Fig.~\ref{fig:wideSqueezed} at the peak of the squeezed bispectrum can be understood as a reflection of other iterative corrections in the GSR hierarchy, modes which converge
		only slowly in this large-step case. 
		These terms are associated
		with the dynamics of the $k_L$ modes and similar corrections are required for
		the power spectrum
		as well.  
		 In fact, it is explicitly shown in Ref.~\cite{Ramirez:2018dxe} 
		that the $g$ terms in the power spectrum expansion reach order unity during the
		transition, which explains why  higher-order GSR contributions are necessary to ensure the consistency relation holds at the bispectrum peak.

		\begin{figure}[t]
		\psfig{file=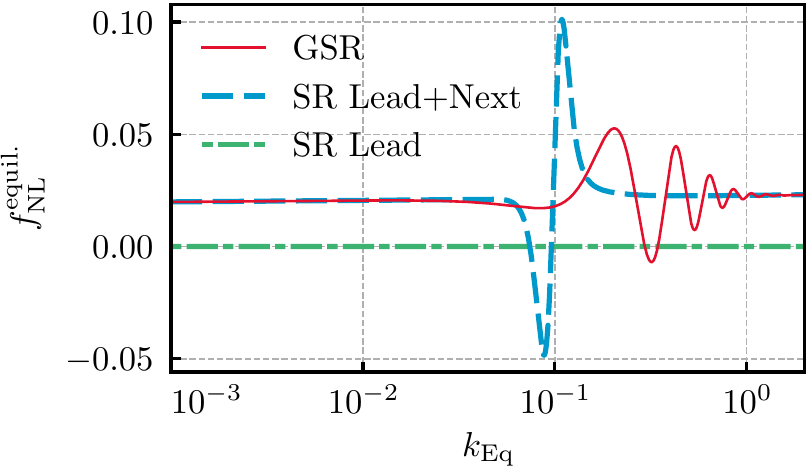}
		\caption{Equilateral bispectrum for small-step transient G-inflation. For this set of parameters, $|f_3| \ll 1$ and thus the leading-order slow-roll contribution Eq.~\eqref{eq:GInfEquilBispectrumSR} remains nearly $0$. The next-to-leading order SR contribution, Eq.~(100) of Ref.~\cite{DeFelice:2013ar}, dominates, and agrees with the GSR computation before and after the step. During the transition, the SR hierarchy is violated and the SR expression fails to accurately track the GSR bispectrum.}
		\label{fig:shortEquil}
		\end{figure}

		\begin{figure}[t]
		\psfig{file=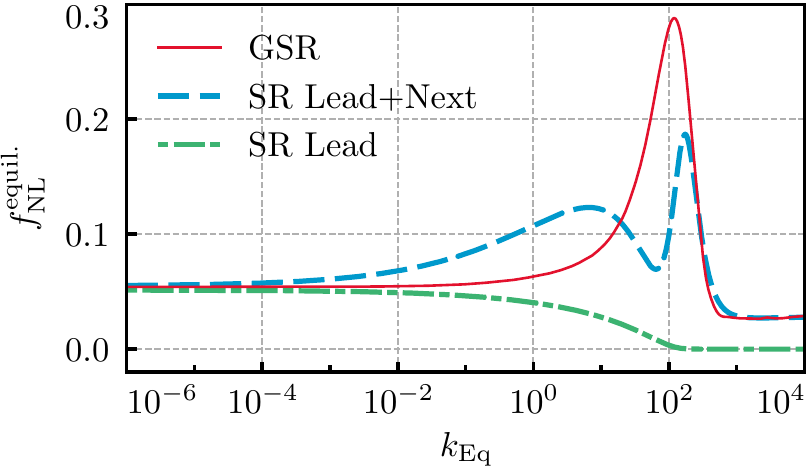}
		\caption{Equilateral bispectrum for large-step transient G-inflation. In this case the leading-order SR contribution  Eq.~\eqref{eq:GInfEquilBispectrumSR} dominates prior to the transition where $|f_3| \gg 1$. After the transition, the leading-order SR contribution goes to zero while the next-to-leading order terms in Eq.~(100) of Ref.~\cite{DeFelice:2013ar} come to dominate. The GSR result again agrees with the SR results before and after the transition, while the SR result shows an erroneous double-peak feature during the transition. Despite the enhancement due to the slow-roll violation, $|f_{\rm NL}^{\rm equil.} | <1$  at all times.}
		\label{fig:wideEquil}
		\end{figure}

	We next turn to the equilateral bispectrum. Only the $i=2$ and $i=3$ operators yield contributions which are not slow-roll suppressed  (see Table~\ref{tab:OperatorSummary}). In the 
	slow-roll approximation one would take their sources to be constant in Eq.~\eqref{eq:GSRBi}
	and obtain 
		\begin{equation}
		\label{eq:EquilBispectrumSR}
		f_{\mathrm{NL}}^{\mathrm{lead;\ equil.}} \simeq \frac{35}{108} (1 - F)_{\textrm{SR}} + \frac{5}{81} \left(\frac{F_3}{Q}\right)_{\textrm{SR}},
		\end{equation}
in which the $``\textrm{SR}"$ subscript denotes that the functions should be expanded to zeroth order in slow-roll. This can be shown to agree analytically with the result for the leading-order equilateral bispectrum in the literature for Horndeski models, Eq.~(97) of Ref.~\cite{DeFelice:2013ar}. In the specific case of transient G-inflation, Eq.~\eqref{eq:EquilBispectrumSR} takes the form

		\begin{align*}
		\label{eq:GInfEquilBispectrumSR}
		&f_{\mathrm{NL}}^{\mathrm{SR, equil.}} \simeq \numberthis \\ &\frac{5 f_3^2 H^2 \dot\phi {}^2 \left(17
		 + 94 f_3 H \dot\phi 
		 - 17 f_{3,\phi} \dot\phi^2\right)}{81 \left(1
		 + 4 f_3 H \dot\phi 
		 - f_{3,\phi} \dot\phi^2\right)^2 \left(1
		 + 6 f_3 H \dot\phi 
		 - f_{3,\phi} \dot\phi^2\right)},
		\end{align*}
		in which ${}_{,\phi}\equiv d/d\phi$.
		
		When $|f_3|$ is large, as in pure G-inflation, the leading-order equilateral bispectrum dominates over slow-roll suppressed terms and leads to a larger bispectrum than in canonical inflation.
		However, when $|f_3|$ is small, as occurs in the small step model and after the transition in the wide step model, the leading-order contribution to the equilateral bispectrum is subdominant to the slow-roll suppressed contributions from the $i=0$ and $i=1$ operators. For this case, Ref.~\cite{DeFelice:2013ar} computes a next-to-leading-order contribution to the bispectrum, which results from considering the contributions from slow-roll suppressed operators, the next order in slow-roll contributions from the $i=2$ and $i=3$ operators, as well as SR corrections to the modefunctions.

		In Fig.~\ref{fig:shortEquil} and Fig.~\ref{fig:wideEquil}, we compare the total equilateral bispectrum in GSR with the leading-order slow-roll expression \eqref{eq:GInfEquilBispectrumSR} as well as Eqs.~(97) and (100) of Ref.~\cite{DeFelice:2013ar}, formulas which include the next-to-leading-order contributions.
		
		In the small-step case, inflation before and after the transition is nearly canonical and thus the equilateral bispectrum is dominated by the $i=0$ operator. At the transition the $i= 1$, $i=2$, and $i= 3$ operators  contribute, while the $i= 4$ and $i= 5$ operators remain subdominant throughout. As expected, the leading-order slow-roll bispectrum is subdominant throughout while the slow-roll formula including the  next-to-leading-order contributions agrees well with GSR before and after the transition. However, during the transition it  displays radically different behavior from the GSR curve and fails to reproduce the oscillatory equilateral bispectrum resulting from the sharp sources.
		
	In the large-step case, inflation before the step is in a G-inflation dominated phase. During this phase, the $i=2$ and $i=3$ operators dominate the equilateral bispectrum. In the G-inflation dominated limit, $f_3 \rightarrow -\infty$, the leading-order contribution in slow-roll to the equilateral bispectrum (\ref{eq:GInfEquilBispectrumSR}) approaches $235/3888 \sim 0.06$. The slow-roll suppressed contribution only yields a small adjustment to this value. This is significantly smaller than might be expected from the $k$-inflation scaling, for example in DBI inflation $f_{\mathrm{NL}}^{\textrm{equil.}} \simeq \frac{35}{108} (1-1/{c_s^2})$, which with the G-inflation $c_s^2 \simeq 2/3$ yields $f_{\mathrm{NL}}^{\textrm{equil.}} \simeq -0.16$.
	Note also the difference in sign.

 After the transition, $f_3 \rightarrow 0$ and the leading-order contribution goes to zero while next-to-leading order contributions become important. Once more, while the leading-order and next-to-leading order SR formulas can accurately track the GSR bispectrum when the usual slow-roll hierarchy is maintained, they fail during the transition when
 this hierarchy is violated. In particular, the next-to-leading order SR formula predicts an erroneous double peak structure in the equilateral bispectrum.

\section{Discussion}
	\label{sec:Discussion}

	In this paper, we develop an effective field theory approach for the study of the bispectrum in single-clock inflation beyond the usual slow-roll approximation. 
	This approach begins with the most general action which breaks temporal diffeomorphisms but preserves spatial diffeomorphisms. 
	In addition we require that the scalar degree of freedom obeys a standard dispersion relation at leading order so that power spectra behave in the usual way. 

	Our approach of studying the action directly in unitary gauge yields a wider set of terms in the action than explicitly considered in previous work \cite{Cheung:2007st,Baumann:2011su,Senatore:2009gt,Bartolo:2010di}, and in particular our action encompasses the Horndeski \cite{Horndeski:1974wa} and GLPV \cite{Gleyzes:2014qga} classes.

	From this starting point we derive the cubic action for scalar curvature perturbations, making use of integration by parts and the equation of motion while discarding boundary terms which are suppressed outside the horizon. 
	By appropriately grouping the operators, we isolate the ones that contribute in the squeezed limit and highlight the consistency relation between the power spectrum and the squeezed bispectrum. 
	The resultant cubic action contains ten operators, of which six are present in the Horndeski and GLPV classes, and of these six operators four are slow-roll suppressed.

	We then compute the tree-level bispectrum contribution for each operator using the in-in and GSR formalisms which are valid beyond the slow-roll limit.
	Our GSR results enable computation of any bispectrum configuration
	for all the operators in our action from a set of simple one-dimensional integrals.
	
	In particular the GSR expressions confirm that the
	 consistency relation holds not just in the slow-roll approximation but also in the case of rapidly
	 varying sources.  
	This result extends works which show that the consistency relation explicitly holds in slow-roll, for specific models, or for certain subclasses of EFT operators \cite{Cheung:2007sv,Bartolo:2013exa,DeFelice:2013ar,Adshead:2013zfa}.

	As an explicit example, we compute the bispectrum for a specific inflationary model in the Horndeski class in which slow-roll is transiently violated, the transient G-inflation model \cite{Ramirez:2018dxe}. For this model, our first-order GSR results for the equilateral bispectrum show qualitatively different behavior from the slow-roll results in the literature during the slow-roll violating phase. This model also highlights corrections for squeezed configurations
	from non-leading GSR terms which can be important in models in which the power spectrum deviates dramatically from scale-invariance between freeze-out epochs. 
	
	The large number of time-dependent coefficients in the EFT of inflation allows a rich range of behavior for the bispectrum beyond slow-roll. By condensing this large family of coefficients  into a small number of integrals, we 
	have provided the tools with which the bispectrum for a very general class of inflation models
	can be easily studied.

\acknowledgments
	We thank H\'ector Ram\'irez and Hayato Motohashi for their assistance with the transient G-inflation model, Andrew J. Long for fruitful discussions, and Peter Adshead for his helpful comments on a draft of this work. This work made use of the MathGR tensor computation package by Yi Wang \cite{Wang:2013mea}.

	SP thanks the Yukawa Institute for Theoretical Physics at Kyoto University, where discussions during the workshop YITP-T-17-02 ``Gravity and Cosmology 2018" were useful in the completion of this work. SP is also grateful to Masahiro Takada and the Kavli Institute for the Physics and Mathematics of the Universe at the University of Tokyo for their warm hospitality.

	SP and WH were supported by U.S.\ Dept.\ of Energy contract DE-FG02-13ER41958, NASA ATP NNX15AK22G and the Simons Foundation.

\onecolumngrid
\appendix
\allowdisplaybreaks
\section{Cubic Action Coefficients}
	\label{app:Fs}

	In this Appendix, we provide the EFT coefficients that appear in the cubic action \eqref{eq:cubicaction}.
	For compactness, we first define some intermediary variables. Prior to temporal integration by parts and equation of motion simplifications, but after spatial integration by parts, the lapse- and shift-eliminated EFT action \eqref{eq:EFTAction} is
	\begin{align*}
	\label{eq:cubiclagrangianjustspatial}
		S_3 =& \int \diffcubed{x} \diff{t} \Bigl[a^3 P_1 \zeta  \dot\zeta^2
		 + a P_2 \zeta  \left(\pa \zeta \right)^2
		 + a^3 P_3 \dot\zeta^3 
		 + a^3 P_4 \dot\zeta  \left(\pa_a\zeta\right) \left(\pa_a\chi\right)
		 + a P_5 \pa^2\zeta  \left(\pa_a\pa_b\chi \right)^2 
		 + \frac{P_6}{a} \dot\zeta  \pa^2\zeta  \pa^2\zeta \\
		 &+ \frac{P_7}{a^3} \pa^2\zeta \left(\pa_a\pa_b\zeta \right)^2 
		 + \frac{P_8}{a^3} \pa^2\zeta  \pa^2\zeta  \pa^2\zeta
		 + \frac{P_9}{a} \pa^2\zeta  \left(\pa_a\pa_b\zeta\right) \left(\pa_a\pa_b\chi\right) 
		 + a P_{10} \dot\zeta {}^2 \pa^2\zeta 
		 + a P_{11} \dot\zeta  \left(\pa \zeta \right)^2\\
		 &+ a P_{12} \zeta  \dot\zeta  \pa^2\zeta 
		 + \frac{P_{13}}{a}\dot\zeta  \left(\pa_a\pa_b\zeta \right)^2
		 + \frac{P_{14}}{a} \left(\pa \zeta \right)^2 \pa^2\zeta 
		 + a P_{15} \dot\zeta  \left(\pa_a\pa_b\zeta\right)\left(\pa_a\pa_b\chi\right)
		 + a P_{16} \pa^2\zeta \left(\pa_a\zeta\right)\left(\pa_a\chi\right)\\
		 &+ a^3 P_{17} \pa^2\zeta  \left(\pa \chi \right)^2 
		 + a^3 P_{18} \dot\zeta  \left(\pa_a\pa_b\chi \right)^2 \Bigr],
	\numberthis
	\end{align*}
	in which
	\begin{align*}
	P_1 =& - 3 \C_{KK} D_3, \\
	P_2 =& \ 2 \C_R,\\
	P_3 =& \ \frac{1}{12} \Bigl[6\C_{\bar{K}\bar{K}K} (1 - H D_1) (9 + 9 H^2 D_1^2 
	 + 9 H D_1 (D_3
	 - 2)
	 - 9 D_3
	 + 2 D_3^2)
	 +\C_{\bar{K}\bar{K}\bar{K}} (6 
	 - 6 H^3 D_1^3 
	 - 6 H^2 D_1^2 (D_3
	 - 3
	 )\\
	 &- 6 D_3
	 + D_3^3 
	 + 6 H D_1 (2 D_3- 3))
	 - 2 (\C_{KKK} (
	 - 3
	 + 3 H D_1
	 + D_3)^3
	 + D_1 (
	 - 27\C_{NKK}
	 + 54 H\C_{NKK} D_1
	 - 9\C_{NNK} D_1\\
	 &- 27 H^2\C_{NKK} D_1^2
	 - 3\C_{NN} D_1^2
	 + 9 H\C_{NNK} D_1^2
	 -\C_{NNN} D_1^2
	 + 18\C_{NKK} D_3
	 - 18 H\C_{NKK} D_1 D_3
	 + 3\C_{NNK} D_1 D_3 \\
	 &- 3\C_{NKK} D_3^2 
	 - 3\C_{N\bar{K}\bar{K}} (
	 H D_1- 1
	 ) (
	 3 H D_1 - 3
	 + 2 D_3)
	 + 3\C_{KK} (6
	 + 6 H^2 D_1^2
	 + 4 H D_1 (
	 D_3- 3
	 )
	 - 4 D_3
	 + D_3^2))) \Bigr],\\
 	P_4 =& \ \frac{1}{2} \C_{KK} D_3^2, \\
 	P_5 =& -\frac{D_3^2}{4} \Bigl[8 \C_{\bar{K}\bar{K}R}
	 + 3 \C_{\bar{K}\bar{K}\bar{R}} 
	 +  2 \C_{\bar{K}\bar{K}K} D_2
	 + \C_{\bar{K}\bar{K}\bar{K}} D_2 \Bigr], \\
	P_6 =& \ \frac{1}{12} \Bigl[192 \C_{\bar{K}\bar{R}R}
	 + 30 \C_{\bar{K}\bar{R}\bar{R}}
	 + 90 \C_{K\bar{R}\bar{R}}
	 + 288 \C_{KRR}
	 - 192 H \C_{\bar{K}\bar{R}R} D_1
	 - 30 H \C_{\bar{K}\bar{R}\bar{R}} D_1
	 - 90 H \C_{K\bar{R}\bar{R}} D_1 \\
	 &- 288 H \C_{KRR} D_1
	 + 30 \C_{N\bar{R}\bar{R}} D_1
	 + 96 \C_{NRR} D_1
	 + 16 \C_{RR} D_1
	 + 48 \C_{\bar{K}\bar{K}R} D_2
	 + 12 \C_{\bar{K}\bar{K}\bar{R}} D_2
	 + 84 \C_{\bar{K}K\bar{R}} D_2\\
	 &+ 144 \C_{KKR} D_2
	 - 48 H \C_{\bar{K}\bar{K}R} D_1 D_2
	 - 12 H \C_{\bar{K}\bar{K}\bar{R}} D_1 D_2 
	 - 84 H \C_{\bar{K}K\bar{R}} D_1 D_2
	 - 144 H \C_{KKR} D_1 D_2
	 + 12 \C_{N\bar{K}\bar{R}} D_1 D_2\\
	 &+ 48 \C_{NKR} D_1 D_2
	 + 12 \C_{\bar{K}\bar{K}K} D_2^2
	 + 18 \C_{KKK} D_2^2
	 - 6 \C_{KK} D_1 D_2^2
	 - 12 H \C_{\bar{K}\bar{K}K} D_1 D_2^2
	 - 18 H \C_{KKK} D_1 D_2^2 \\
	 &+ 6 \C_{NKK} D_1 D_2^2
	 - 48 \C_{\bar{K}\bar{R}R} D_3
	 - 3 \C_{\bar{K}\bar{R}\bar{R}} D_3
	 - 30 \C_{K\bar{R}\bar{R}} D_3
	 - 96 \C_{KRR} D_3
	 + 6 \C_{\bar{K}\bar{K}\bar{R}} D_2 D_3
	 - 24 \C_{\bar{K}K\bar{R}} D_2 D_3\\
	 &- 48 \C_{KKR} D_2 D_3
	 + 3 \C_{\bar{K}\bar{K}\bar{K}} D_2^2 D_3
	 - 6 \C_{KKK} D_2^2 D_3 \Bigr], \\
	P_7 =& \ \frac{1}{4}\Bigl[- 8\C_{\bar{R}\bar{R}R} - 3\C_{\bar{R}\bar{R}\bar{R}}
	 - 16\C_{\bar{K}\bar{R}R} D_2 
	 - 7\C_{\bar{K}\bar{R}\bar{R}} D_2
	 - 2\C_{K\bar{R}\bar{R}} D_2
	 - 8\C_{\bar{K}\bar{K}R} D_2^2 
	 - 5\C_{\bar{K}\bar{K}\bar{R}} D_2^2
	 - 4\C_{\bar{K}K\bar{R}} D_2^2\\
	 &- 2\C_{\bar{K}\bar{K}K} D_2^3  
	 -  \C_{\bar{K}\bar{K}\bar{K}} D_2^3\Bigr], \\
	P_8 =& -\frac{1}{12} \Bigl[120\C_{\bar{R}\bar{R}R}
	 + 11\C_{\bar{R}\bar{R}\bar{R}}
	 + 128\C_{RRR} 
	 + 48\C_{\bar{K}\bar{R}R} D_2
	 + 3\C_{\bar{K}\bar{R}\bar{R}} D_2
	 + 30\C_{K\bar{R}\bar{R}} D_2 
	 + 96\C_{KRR} D_2
	 - 3\C_{\bar{K}\bar{K}\bar{R}} D_2^2 \\
	 &+ 12\C_{\bar{K}K\bar{R}} D_2^2 
	 + 24\C_{KKR} D_2^2
	 -\C_{\bar{K}\bar{K}\bar{K}} D_2^3
	 +  2\C_{KKK} D_2^3\Bigr], \\ 
 	P_9 =& -\frac{D_3}{2} \Bigl[8\C_{\bar{K}\bar{R}R}
	 + 3\C_{\bar{K}\bar{R}\bar{R}}
	 + 8\C_{\bar{K}\bar{K}R} D_2  
	 + 4\C_{\bar{K}\bar{K}\bar{R}} D_2
	 + 2\C_{\bar{K}K\bar{R}} D_2
	 + 2\C_{\bar{K}\bar{K}K} D_2^2
	 + \C_{\bar{K}\bar{K}\bar{K}} D_2^2 \Bigr],\\
	P_{10} =& \ \frac{1}{4} \Bigl[
 - 24 \C_{\bar{K}\bar{K}R}
 - 8 \C_{\bar{K}\bar{K}\bar{R}}
 - 48 \C_{\bar{K}K\bar{R}}
 - 72 \C_{KKR}
 + 48 H \C_{\bar{K}\bar{K}R} D_1
 + 16 H \C_{\bar{K}\bar{K}\bar{R}} D_1
 + 96 H \C_{\bar{K}K\bar{R}} D_1
 + 144 H \C_{KKR} D_1 \\
 &- 16 \C_{N\bar{K}\bar{R}} D_1
 - 48 \C_{NKR} D_1 
 - 24 H^2 \C_{\bar{K}\bar{K}R} D_1^2
 - 8 H^2 \C_{\bar{K}\bar{K}\bar{R}} D_1^2
 - 48 H^2 \C_{\bar{K}K\bar{R}} D_1^2
 - 72 H^2 \C_{KKR} D_1^2
 + 16 H \C_{N\bar{K}\bar{R}} D_1^2\\
 &+ 48 H \C_{NKR} D_1^2
 - 8 \C_{NNR} D_1^2
 - 16 \C_{NR} D_1^2
 - 18 \C_{\bar{K}\bar{K}K} D_2
 - 2 \C_{\bar{K}\bar{K}\bar{K}} D_2
 - 18 \C_{KKK} D_2
 + 8 \C_{KK} D_1 D_2\\
 &+ 36 H \C_{\bar{K}\bar{K}K} D_1 D_2
 + 4 H \C_{\bar{K}\bar{K}\bar{K}} D_1 D_2
 + 36 H \C_{KKK} D_1 D_2 
 - 4 \C_{N\bar{K}\bar{K}} D_1 D_2
 - 12 \C_{NKK} D_1 D_2
 - 8 H \C_{KK} D_1^2 D_2 \\
 &- 18 H^2 \C_{\bar{K}\bar{K}K} D_1^2 D_2
 - 2 H^2 \C_{\bar{K}\bar{K}\bar{K}} D_1^2 D_2
 - 18 H^2 \C_{KKK} D_1^2 D_2
 + 4 H \C_{N\bar{K}\bar{K}} D_1^2 D_2
 + 12 H \C_{NKK} D_1^2 D_2\\
 &- 2 \C_{NNK} D_1^2 D_2
 + 16 \C_{\bar{K}\bar{K}R} D_3
 + 4 \C_{\bar{K}\bar{K}\bar{R}} D_3
 + 28 \C_{\bar{K}K\bar{R}} D_3
 + 48 \C_{KKR} D_3
 - 16 H \C_{\bar{K}\bar{K}R} D_1 D_3
 - 4 H \C_{\bar{K}\bar{K}\bar{R}} D_1 D_3\\
 &- 28 H \C_{\bar{K}K\bar{R}} D_1 D_3
 - 48 H \C_{KKR} D_1 D_3
 + 4 \C_{N\bar{K}\bar{R}} D_1 D_3
 + 16 \C_{NKR} D_1 D_3
 + 8 \C_{\bar{K}\bar{K}K} D_2 D_3
 + 12 \C_{KKK} D_2 D_3\\
 &- 4 \C_{KK} D_1 D_2 D_3
 - 8 H \C_{\bar{K}\bar{K}K} D_1 D_2 D_3
 - 12 H \C_{KKK} D_1 D_2 D_3
 + 4 \C_{NKK} D_1 D_2 D_3
 + \C_{\bar{K}\bar{K}\bar{R}} D_3^2
 - 4 \C_{\bar{K}K\bar{R}} D_3^2\\
 &- 8 \C_{KKR} D_3^2
 + \C_{\bar{K}\bar{K}\bar{K}} D_2 D_3^2
	 - 2 \C_{KKK} D_2 D_3^2\Bigr], \\
	P_{11} =& \ \frac{1}{2} \Bigl[- 4\C_{KR}
	 + 4 H\C_{KR} D_1
	 - 4\C_{NR} D_1 
	 - 4\C_{R} D_1
	 - 4\C_{KK} D_2
	 + 4 H\C_{KK} D_1 D_2 
	 -  2\C_{NK} D_1 D_2
	 +\C_{KK} D_2 D_3\Bigr], \\
	  P_{12} =& \ \frac{4\C_{KR}\C_{NK}
	 - 8\C_{KK} \left(\C_{NR}
	 +\C_{R}\right)}{2 H\C_{KK}
	 -\C_{NK}}, \\
	 P_{13} =& \ \frac{1}{12} \Bigl[6\C_{N\bar{R}\bar{R}} D_1
	 - 16\C_{RR} D_1
	 + 12\C_{\bar{K}\bar{K}\bar{R}} D_2  
	 + 36\C_{\bar{K}K\bar{R}} D_2
	 - 12 H\C_{\bar{K}\bar{K}\bar{R}} D_1 D_2 
	 - 36 H\C_{\bar{K}K\bar{R}} D_1 D_2
	 + 12\C_{N\bar{K}\bar{R}} D_1 D_2 \\
	 &+ 18\C_{\bar{K}\bar{K}K} D_2^2
	 + 6\C_{\bar{K}\bar{K}\bar{K}} D_2^2
	 + 6\C_{KK} D_1 D_2^2 
	 - 18 H\C_{\bar{K}\bar{K}K} D_1 D_2^2
	 - 6 H\C_{\bar{K}\bar{K}\bar{K}} D_1 D_2^2 
	 + 6\C_{N\bar{K}\bar{K}} D_1 D_2^2\\
	 &- 6\C_{\bar{K}\bar{K}\bar{R}} D_2 D_3
	 - 12\C_{\bar{K}K\bar{R}} D_2 D_3 
	 - 6\C_{\bar{K}\bar{K}K} D_2^2 D_3
	 - 3\C_{\bar{K}\bar{K}\bar{K}} D_2^2 D_3 
	 - 3\C_{\bar{K}\bar{R}\bar{R}} \left(
	 - 2
	 + 2 H D_1
	 + D_3\right) \\
	 &- 6\C_{K\bar{R}\bar{R}} \left(
	 - 3
	 + 3 H D_1
	 + D_3\right) \Bigr], \\
	  P_{14} =& \ \frac{1}{12} \Bigl[8\C_{RR}
	 - 3\C_{KK} D_2^2\Bigr], \\
	 P_{15} =& -\frac{D_3}{2} \Bigl[
	 - 2\C_{N\bar{K}\bar{R}} D_1
	 - 6\C_{\bar{K}\bar{K}K} D_2
	 - 2\C_{\bar{K}\bar{K}\bar{K}} D_2  
	 - 2\C_{KK} D_1 D_2
	 + 6 H\C_{\bar{K}\bar{K}K} D_1 D_2
	 + 2 H\C_{\bar{K}\bar{K}\bar{K}} D_1 D_2 \\
	 &- 2\C_{N\bar{K}\bar{K}} D_1 D_2
	 + 2\C_{\bar{K}\bar{K}K} D_2 D_3
	 +\C_{\bar{K}\bar{K}\bar{K}} D_2 D_3 
	 +\C_{\bar{K}\bar{K}\bar{R}} \left(
	 - 2
	 + 2 H D_1
	 + D_3\right) 
	 + 2\C_{\bar{K}K\bar{R}} \left(
	 - 3
	 + 3 H D_1
	 + D_3\right)  \Bigr], \\
		P_{16} =& -\C_{KK} D_2 D_3, \\
		P_{17} =& -\frac{3}{4}\C_{KK} D_3^2, \\
		P_{18} =& -\frac{1}{4} D_3^2 \Bigl[
	     - 2 \left(\C_{KK}
		 +\C_{N\bar{K}\bar{K}}\right) D_1  
		 +\C_{\bar{K}\bar{K}\bar{K}} \left(
		 - 2
		 + 2 H D_1
		 + D_3\right) 
		 + 2\C_{\bar{K}\bar{K}K} \left(
		 - 3
		 + 3 H D_1
		 + D_3\right)\Bigr] ,
	\numberthis
	\end{align*}
		where $D_1$, $D_2$, and $D_3$ were defined in Eq.~\eqref{eq:Dn}.

 	The cubic action \eqref{eq:cubiclagrangiandirect} then results from performing a series of time integrations by parts and equation of motion simplifications on Eq.~\eqref{eq:cubiclagrangianjustspatial}. The resulting operator coefficients are
	
	\begin{align*}
	F_1 =& \ P_1 -\frac{H}{2 c_s^4} \Bigl(2 c_s^2 H \left(30
	 + 10 q^2
	 - 5 q_2
	 + 10 \e 
	 + 26 \sigma
	 + 2 \e  \sigma
	 + 4 \sigma^2
	 + q (35
	 + 5 \e 
	 + 14 \sigma)
	 - 2 \sigma_2\right) P_5
	 \\&+ 2 c_s^2 (3
	 + 2 q
	 + 2 \sigma) P_{11}
	 - 3 c_s^2 P_{12}
	 - 2 c_s^2 q P_{12}
	 - 2 c_s^2 \sigma P_{12}
	 + 6 H^2 P_{13}
	 + 7 H^2 q P_{13}
	 + 2 H^2 q^2 P_{13}
	 - H^2 q_2 P_{13}
	 \\&+ 7 H^2 \e  P_{13}
	 + 4 H^2 q \e  P_{13}
	 - 2 H^2 \delta_1 \e  P_{13}
	 + 14 H^2 \sigma P_{13}
	 + 8 H^2 q \sigma P_{13}
	 + 8 H^2 \e  \sigma P_{13}
	 + 8 H^2 \sigma^2 P_{13}
	 - 2 H^2 \sigma_2 P_{13}
	 \\&+ 12 H P_{14}
	 + 14 H q P_{14}
	 + 4 H q^2 P_{14}
	 - 2 H q_2 P_{14}
	 + 4 H \e  P_{14}
	 + 2 H q \e  P_{14}
	 + 28 H \sigma P_{14}
	 + 16 H q \sigma P_{14}
	 + 4 H \e  \sigma P_{14}
	 \\&+ 16 H \sigma^2 P_{14}
	 - 4 H \sigma_2 P_{14}
	 - 12 c_s^2 H P_{15}
	 - 14 c_s^2 H q P_{15}
	 - 4 c_s^2 H q^2 P_{15}
	 + 2 c_s^2 H q_2 P_{15}
	 - 4 c_s^2 H \e  P_{15}
	 - 2 c_s^2 H q \e  P_{15}
	 \\&- 8 c_s^2 H \sigma P_{15}
	 - 4 c_s^2 H q \sigma P_{15}
	 - 6 c_s^2 P_{16}
	 - 4 c_s^2 q P_{16}
	 - 4 c_s^2 \sigma P_{16}
	 + 6 c_s^4 P_{18}
	 + 4 c_s^4 q P_{18}
	 - 38 c_s^2 H p_5
	 - 22 c_s^2 H q p_5
	 \\&- 6 c_s^2 H \e  p_5
	 - 16 c_s^2 H \sigma p_5
	 - 2 c_s^2 p_{11}
	 + c_s^2 p_{12}
	 - 11 H^2 p_{13}
	 - 10 H^2 q p_{13}
	 - 2 H^2 q^2 p_{13}
	 + H^2 q_2 p_{13}
	 - 10 H^2 \e  p_{13}
	 \\&- 4 H^2 q \e  p_{13}
	 + 2 H^2 \delta_1 \e  p_{13}
	 - 20 H^2 \sigma p_{13}
	 - 8 H^2 q \sigma p_{13}
	 - 8 H^2 \e  \sigma p_{13}
	 - 8 H^2 \sigma^2 p_{13}
	 + 2 H^2 \sigma_2 p_{13}
	 - 10 H p_{14}
	 \\&- 6 H q p_{14}
	 - 2 H \e  p_{14}
	 - 12 H \sigma p_{14}
	 + 10 c_s^2 H p_{15}
	 + 6 c_s^2 H q p_{15}
	 + 2 c_s^2 H \e  p_{15}
	 + 4 c_s^2 H \sigma p_{15}
	 + 2 c_s^2 p_{16}
	 - 2 c_s^4 p_{18}
	 \\&+ 6 c_s^2 H p_{5,2}
	 + 6 H^2 p_{13,2}
	 + 3 H^2 q p_{13,2}
	 + 3 H^2 \e  p_{13,2}
	 + 6 H^2 \sigma p_{13,2}
	 + 2 H p_{14,2}
	 - 2 c_s^2 H p_{15,2}
	 - H^2 p_{13,3}\Bigr),\\
	 F_2 =& \ P_2 + \frac{H}{2 c_s^2} \Bigl(2 c_s^2 H (5 q_2
	 - 5 q (
	 - 1
	 + \e )
	 + 2 (5
	 + \sigma
	 - \e  (5
	 + \sigma)
	 + \sigma_2)) P_5
	 + 2 c_s^2 P_{11}
	 + 2 H^2 P_{13}
	 + H^2 q P_{13}
	 + H^2 q_2 P_{13}
	 \\&- 3 H^2 \e  P_{13}
	 - 2 H^2 q \e  P_{13}
	 + 2 H^2 \delta_1 \e  P_{13}
	 - 2 H^2 \sigma P_{13}
	 - 2 H^2 q \sigma P_{13}
	 - 6 H^2 \e  \sigma P_{13}
	 - 4 H^2 \sigma^2 P_{13}
	 + 2 H^2 \sigma_2 P_{13}
	 \\&+ 4 H P_{14}
	 + 2 H q P_{14}
	 + 2 H q_2 P_{14}
	 - 4 H \e  P_{14}
	 - 2 H q \e  P_{14}
	 - 4 H \sigma P_{14}
	 - 4 H q \sigma P_{14}
	 - 4 H \e  \sigma P_{14}
	 - 8 H \sigma^2 P_{14}
	 \\&+ 4 H \sigma_2 P_{14}
	 - 4 c_s^2 H P_{15}
	 - 2 c_s^2 H q P_{15}
	 - 2 c_s^2 H q_2 P_{15}
	 + 4 c_s^2 H \e  P_{15}
	 + 2 c_s^2 H q \e  P_{15}
	 - 2 c_s^2 P_{16}
	 + 2 c_s^4 P_{18}
	 + 4 c_s^4 \sigma P_{18}
	 \\&+ 14 c_s^2 H p_5
	 + 10 c_s^2 H q p_5
	 + 6 c_s^2 H \e  p_5
	 + 4 c_s^2 H \sigma p_5
	 + 2 c_s^2 p_{11}
	 - H^2 p_{13}
	 - H^2 q_2 p_{13}
	 + 6 H^2 \e  p_{13}
	 + 2 H^2 q \e  p_{13}
	 \\&- 2 H^2 \delta_1 \e  p_{13}
	 + 6 H^2 \sigma p_{13}
	 + 2 H^2 q \sigma p_{13}
	 + 6 H^2 \e  \sigma p_{13}
	 + 4 H^2 \sigma^2 p_{13}
	 - 2 H^2 \sigma_2 p_{13}
	 + 2 H p_{14}
	 + 2 H q p_{14}
	 + 2 H \e  p_{14}
	 \\&+ 8 H \sigma p_{14}
	 - 2 c_s^2 H p_{15}
	 - 2 c_s^2 H q p_{15}
	 - 2 c_s^2 H \e  p_{15}
	 - 2 c_s^2 p_{16}
	 + 2 c_s^4 p_{18}
	 - 6 c_s^2 H p_{5,2}
	 - 2 H^2 p_{13,2}
	 - H^2 q p_{13,2}
	 \\&- 3 H^2 \e  p_{13,2}
	 - 4 H^2 \sigma p_{13,2}
	 - 2 H p_{14,2}
	 + 2 c_s^2 H p_{15,2}
	 + H^2 p_{13,3}\Bigr),
	\\F_3 =& \ {H} P_3 + \frac{{H}}{6 c_s^4} \Bigl(48 c_s^2 H P_5
	 + 24 c_s^2 H q P_5
	 + 10 c_s^2 H \sigma P_5
	 + 12 c_s^2 H P_{10}
	 + 6 c_s^2 H q P_{10}
	 + 4 c_s^2 H \sigma P_{10}
	 + 6 c_s^2 P_{11}
	 - 3 c_s^2 P_{12}
	 \\&+ 12 H^2 P_{13}
	 + 6 H^2 q P_{13}
	 + 4 H^2 \e  P_{13}
	 + 10 H^2 \sigma P_{13}
	 + 24 H P_{14}
	 + 12 H q P_{14}
	 + 20 H \sigma P_{14}
	 - 6 c_s^2 H P_{15}
	 - 3 c_s^2 H q P_{15}
	 \\&+ c_s^2 H \sigma P_{15}
	 - 3 c_s^2 P_{16}
	 + 6 c_s^4 P_{18}
	 - 14 c_s^2 H p_5
	 - 2 c_s^2 H p_{10}
	 - 16 H^2 p_{13}
	 - 6 H^2 q p_{13}
	 - 4 H^2 \e  p_{13}
	 - 10 H^2 \sigma p_{13}
	 - 8 H p_{14}
	 \\&+ 4 c_s^2 H p_{15}
	 + 4 H^2 p_{13,2} \Bigr),\\
	F_4 =& \ P_4 - \frac{H}{2 c_s^2} \Bigl(48 H P_5
	 + 56 H q P_5
	 + 16 H q^2 P_5
	 - 8 H q_2 P_5
	 + 16 H \e  P_5
	 + 8 H q \e  P_5
	 + 44 H \sigma P_5
	 + 24 H q \sigma P_5
	 + 4 H \e  \sigma P_5 
	 \\&+ 8 H \sigma^2 P_5
	 - 4 H \sigma_2 P_5
	 - 6 H P_{15}
	 - 7 H q P_{15}
	 - 2 H q^2 P_{15}
	 + H q_2 P_{15}
	 - 2 H \e  P_{15}
	 - H q \e  P_{15}
	 - 4 H \sigma P_{15}
	 - 2 H q \sigma P_{15}
	 \\&- 6 P_{16}
	 - 4 q P_{16}
	 - 4 \sigma P_{16}
	 + 6 c_s^2 P_{18}
	 + 4 c_s^2 q P_{18}
	 - 28 H p_5
	 - 16 H q p_5
	 - 4 H \e  p_5
	 - 12 H \sigma p_5
	 + 5 H p_{15}
	 + 3 H q p_{15}
	 \\&+ H \e  p_{15}
	 + 2 H \sigma p_{15}
	 + 2 p_{16}
	 - 2 c_s^2 p_{18}
	 + 4 H p_{5,2}
	 - H p_{15,2}\Bigr),\\
	F_5 =& \ P_{17} + \frac{d}{dt} \Bigl(\frac{H}{ 2 c_s^2} \bigl(
	 p_5- P_5(6
	 + 3 q
	 + 2 \sigma)
	 \bigr)\Bigr)
	 + \frac{H^2 P_5}{2 c_s^2} (3
	 + 2 q) (6
	 + 3 q
	 + 2 \sigma)
	 - \frac{H}{2} \bigl( p_{18}
	 -  P_{18}(3
	 + 2 q)
	 + \frac{H p_5}{c_s^2} (3
	 + 2 q)
	 \bigr),\\
	F_6 =& \ {H^3} (P_6 + P_{13}), \\
	F_7 =& \ {H^4} P_7, \\
	F_8 =& \ {H^4} P_8, \\
	F_9 =& \ {H^3} P_9,
	\numberthis
	\end{align*}
	in which  $p_{n,k} \equiv d p_{n,k-1}/ dN$ and $p_{n,1} \equiv p_n \equiv d P_n / dN$;
	$\sigma_{k} \equiv d \sigma_{k-1}/ dN$ and $\sigma_{1} \equiv \sigma$;
	$q_{k} \equiv d q_{k-1}/ dN$ and $q_{1} \equiv q$; and
	$\delta_1 \equiv (1/2)(d\ln \e)/(dN) - \e$.

\section{ Bispectrum Sources, Windows and Configuration Weights}
	\label{app:EFTIntegrals}

	In this Appendix, we provide the full set of sources, windows, and configuration weights which make up Eq.~\eqref{eq:GSRBi}, the GSR integral formulation for the bispectrum.

	\subsection*{\texorpdfstring{$\bm{i=0}$}{i=0}: \texorpdfstring{$\zeta^2 \dot{\zeta}$}{} and \texorpdfstring{$\bm{i=1}$}{i=1}: \texorpdfstring{$\zeta  (\Ha_2 + \L_2)$}{}}

		These operators correspond to the interaction Hamiltonians
		\begin{align*}
		H_I^{i=0} &= - \int \diffcubed{x} \ a^3 Q \frac{\diff{}}{\diff{t}} \left(\e + \frac{3}{2}\sigma + \frac{q}{2}\right) \zeta^2 \dot{\zeta} + \int \diffcubed{x} \frac{\diff{}}{\diff{t}} \left[ a^3 Q \left(\e + \frac{3}{2}\sigma + \frac{q}{2}\right) \zeta^2 \dot{\zeta}\right], \\
		H_I^{i=1} &= - \int \diffcubed{x} (\sigma + \e) \zeta(\Ha_2 + 2 \L_2)
		\numberthis
		\end{align*}
and as we shall see are slow-roll suppressed. Therefore the zeroth-order GSR modefunction result is first order in slow-roll parameters. These operators are considered to the desired GSR order in the context of canonical and $k$-inflation theories in Refs.~\cite{Adshead:2011bw,Adshead:2013zfa}. The results there hold with the source substitutions
		\begin{equation}
		S_{00} = S_{01} = S_{02} = \frac{\left(2 \e + 3 \sigma + q\right)}{2f}, \qquad \qquad
		S_{10} = S_{11} = S_{12} = \frac{\sigma+\e}{f}. 
		\end{equation}
		These sources have corresponding windows
		\begin{equation}
		W_{00} = W_{10} = \ x \sin x, \qquad
		W_{01} = W_{11} = \ \cos x, \qquad 
		W_{02} = W_{12} = \ \frac{\sin x}{x},
		\end{equation}
		and $k$-weights
		\begin{align*}
		&T_{00} = -1, \quad	
		&T_{01} = - \frac{ \sum_{i\ne j} k_i k_j^2}{ k_1 k_2 k_3} , \quad
		&&T_{02} = \frac{K\sum_i k_i^2}{k_1k_2k_3} , 
		\end{align*}
		\begin{align*}
		&T_{10} = \ \frac{3}{2} - \frac{\sum_i k_i^2}{K^2}, \quad 
&T_{11} = \frac{1}{k_1 k_2 k_3} \Big[  \frac{1}{2} \sum_{i\ne j} k_i k_j^2  + \frac{4}{ K} \sum_{i>j} k_i^2 k_j^2 - \frac{2}{K^2} \sum_{i\ne j} k_i^2 k_j^3  
\Big] , \quad 
&&T_{12} =  -\frac{K \sum_i k_i^2}{2 k_1 k_2 k_3}. 
\numberthis
		\end{align*}

		These operators enforce the squeezed-limit consistency relation, as shown in detail in \S\ref{subsec:squeeze}.

	\subsection*{\texorpdfstring{$\bm{i=2}$}{i=2}: \texorpdfstring{$\dot{\zeta} \L_2$}{} and \texorpdfstring{$\bm{i=3}$}{i=3}: \texorpdfstring{$\dot{\zeta}^3$}{}}

		These operators correspond to the interaction Hamiltonians
		\begin{equation}
		H_I^{i=2} = - \int \diffcubed{x} \ (1 - F) \frac{\dot{\zeta} \L_2}{H}, \qquad  \qquad
		H_I^{i=3} = - \int \diffcubed{x} \ a^3 \frac{F_3}{{H}} \dot\zeta {}^3.
		\numberthis
		\end{equation}

		These operators, and all subsequent operators, are not slow-roll suppressed in the EFT. These operators are also not slow-roll suppressed in the $k$-inflation or Horndeski subclasses. The first-order GSR result therefore requires computing the contributions from first-order modefunction corrections, which is done in Ref.~\cite{Adshead:2013zfa}. The results there hold with the source substitutions
		\begin{equation}
		S_2 = \frac{c_s}{a H s} \frac{(F-1)}{f}, \qquad \qquad
		S_3 = - \frac{1}{Q} \frac{c_s}{a {H} s} \frac{F_3}{f}.
		\end{equation}

		In particular, the contributions which are first order in GSR due to the time variation of the sources have the sources
		\begin{align*}
		S_{20} =& S_{21} = S_2' , && S_{30} = S_{31} = S_3' ,\\
		S_{22} =& S_2, && 		S_{32} = S_3, \numberthis
		\end{align*}
		with the windows
		\begin{equation}
		W_{20} = W_{30} =  x \sin x, \qquad \qquad W_{21} = W_{22} = W_{31} = W_{32} = \cos x, 
		\end{equation}
		and the $k$-weights
		\begin{align*}
		T_{20} 	=& \frac{\sum_i k_i^2 - 2 \sum_{i>j} k_i k_j}{2 K^2}, \qquad 
		T_{21}  = \frac{1}{k_1 k_2 k_3} \Big[  \frac{1}{2} \sum_{i\ne j} k_i k_j^2  - \frac{6}{ K} \sum_{i>j} k_i^2 k_j^2 + \frac{4}{K^2} \sum_{i\ne j} k_i^2 k_j^3  
		\Big] - \frac{1}{2} ,\\
		T_{22} =& \frac{1}{k_1 k_2 k_3} \Big[ \frac{1}{2}\sum_i k_i^3 - \frac{4}{K} \sum_{i>j} k_i^2 k_j^2 + \frac{2}{ K^2} \sum_{i\ne j} k_i^2 k_j^3  \Big], \qquad	T_{30} = \frac{T_{31}}{3} = \frac{T_{32}}{2} = -\frac{3k_1 k_2 k_3}{K^3},
		\numberthis
		\end{align*}
		while the contributions which are first order in GSR due to the time variation of the source have the sources
		\begin{equation}
		g S_2 = S_{23}'=S_{24}'=S_{25}'=S_{26}'  , \qquad \qquad 
		g S_3 = S_{33}'=S_{34}'=S_{35}',
		\end{equation}
		the windows
		\begin{align*}
		W_{23} &=  W_{33} = x \sin x + \cos x, 
		&&W_{24} = \cos x, \\
		W_{25} &=  W_{34} = 2\frac{\sin x}{x} - \cos x, 
		&&W_{26} = W_{35} = W_{2B} = W_{3B} =  12\left(  \frac{\sin x}{x^3} - \frac{\cos x}{x^2} -\frac{\sin x}{ 4 x}\right),
		\numberthis
		\end{align*}
		and the $k$-weights
		\begin{align*}
		T_{23} =& \ \frac{3}{2} -\frac{2\sum_{i>j} k_i k_j }{K^2}, \qquad T_{24} = \ \frac{(K-2k_1)(K - 2k_2)(K-2k_3)}{4 k_1 k_2 k_3}, \\
		T_{25} =& \ -\frac{1}{8 k_1 k_2 k_3 (K-2k_3) K^2} \Big[(k_1^2-k_2^2)^2 (k_1^2 + 6 k_1 k_2 + k_2^2) + 4 (k_1-k_2)^2 (k_1+k_2) (k_1^2 + 6 k_1 k_2 + k_2^2) k_3  \\
		 &+ 2 (3 k_1^4 + 23 k_1^3 k_2 + 64 k_1^2 k_2^2 + 23 k_1 k_2^3 + 3 k_2^4)k_3^2 
		 + 16 k_1 k_2 (k_1 + k_2) k_3^3 
		 - (7 k_1^2 + 20 k_1 k_2 + 7 k_2^2) k_3^4  \\
		 &- 4(k_1 + k_2) k_3^5 \Big] + {\rm perm.}, \\
		T_{26} =& \ \frac{1}{12 k_1 k_2 k_3 (K-2k_3)^2 K^2}  \Big[ 
		 (k_1-k_2)^2(k_1+k_2)^3 (k_1^2+ 6 k_1 k_2 + k_2^2) 
		+3(k_1^2-k_2^2)^2 (k_1^2 + 6 k_1 k_2 + k_2^2) k_3  \\
		&+ 2(k_1+k_2)(6 k_1^4 + 35 k_1^3 k_2 + 106 k_1^2 k_2^2
		+ 35 k_1 k_2^3 + 6 k_2^4) k_3^2
		+ 2(2 k_1^4 + 5 k_1^3 k_2 - 26 k_1^2 k_2^2 + 5 k_1 k_2^3 + 2 k_2^4) k_3^3 \\
		&-(k_1 + k_2)(19 k_1^2 + 44 k_1 k_2 + 19 k_2^2) k_3^4 
		+(-9 k_1^2 + 4 k_1 k_2 - 9 k_2^2) k_3^5
		+ 6 (k_1 + k_2) k_3^6 + 2 k_3^7 \Big] + {\rm perm.}, \\
		T_{2B} =& \frac{1}{6k_1 k_2 k_3 (K-2 k_3)^2}
		 \Big[
		( k_1-k_2)^2 (k_1 + k_2) (k_1^2 + 3 k_1 k_2 + k_2^2) 
		- 2 (k_1-k_2)^2 ( k_1^2 + 3 k_1 k_2 + k_2^2) k_3 
		\\&- (k_1 + k_2) (3 k_1^2 + 5 k_1 k_2 + 3 k_2^2) k_3^2 
		+ (3 k_1^2 - 2 k_1 k_2 + 3 k_2^2) k_3^3 
		+ 2 (k_1+k_2) k_3^4 - k_3^5 \Big], \\
		T_{33} =& \frac{3 k_1 k_2 k_3}{K^2 (K- 2 k_3)} +{\rm perm.}, \qquad T_{34} = - \frac{3 k_1 k_2 k_3}{K^2 (K- 2 k_3)^2} \left[ 7(k_1+k_2)-3 k_3 \right] +{\rm perm.},
		\\
		T_{35} =& \frac{4 k_1 k_2 k_3}{K^2 (K-2 k_3)^3} \left[ 5 (k_1+k_2)^2 - 5(k_1+k_2) k_3 + 2 k_3^2 \right] + {\rm perm.}, \quad
		T_{3B} = -\frac{2  k_1 k_2 k_3}{(K-2 k_3)^3}.
		\numberthis
		\end{align*}

		The $k$-weights satisfy the relations
		\begin{equation}
		\sum_{j=3}^{6} T_{2j} + \left[T_{2B} + \text{perm.}\right] = 0, \qquad \qquad \sum_{j=3}^{5} T_{3j} + \left[T_{3B} + \text{perm.}\right] = 0,
		\end{equation}
		which ensures that the outside-the-horizon boundary terms in $I_{2j}$ and $I_{3j}$ for these contributions cancel and therefore that $g S_2$ and $g S_3$ do not need to be integrated. 

		In the squeezed-limit $x_L / x_S \gg 1$, $x_S \ll 1$, these operators satisfy
		\begin{equation}
		\sum_{j} T_{2j} I_{2j} + \left[T_{2B} I_{2B}(2 k_3) + \text{perm.}\right] = 0, \qquad \qquad
		\sum_{j} T_{3j} I_{3j} + \left[T_{3B} I_{3B}(2 k_3) + \text{perm.}\right] = 0,
		\end{equation}
		and therefore these operators have no squeezed contribution.

	\subsection*{\texorpdfstring{$\bm{i=4}$}{i=4}: \texorpdfstring{$\dot{\zeta} \pa_a \zeta \pa_a \chi$}{}}

		This operator corresponds to the interaction Hamiltonian
		\begin{equation}
		H_I^{i=4} =  -\int \diffcubed{x} \ a^3 F_{4} \dot{\zeta} \pa \zeta \pa \chi.
		\end{equation}

		This operator is considered in Ref.~\cite{Adshead:2013zfa} to zeroth order in GSR modefunctions because it is slow-roll suppressed in $k$-inflation, Horndeski, and GLPV models. It is not necessarily slow-roll suppressed in the general EFT. Therefore we present it here to first order in GSR modefunctions.

		We define the source
		\begin{equation}
		S_4 \equiv -\frac{1}{2Q}\frac{F_4}{f},
		\end{equation}
		in which the factor of $2$ here facilitates comparison with Ref.~\cite{Adshead:2013zfa}.

		The contributions which are first order in GSR due to the time variation of the source have the sources
		\begin{equation}
		S_{40} = S_{41} = S_4,
		\end{equation}
		the windows
		\begin{equation}
		W_{40} = x \sin{x}, \quad W_{41} = \cos{x},
		\end{equation}
		and the $k$-weights 
		\begin{equation}
		T_{40} = -\frac{1}{K^2} \frac{1}{k_1 k_2 k_3} \Bigl[(\mathbf{k_1}\cdot \mathbf{k_2}) k_3^2 (k_1 + k_2)+\text{perm.}\Bigr], \quad 
		T_{41} = -\frac{1}{K^2} \frac{1}{k_1 k_2 k_3} \Bigl[(\mathbf{k_1}\cdot \mathbf{k_2}) k_3^2 (3 K - k_3)+ \text{perm.}\Bigr].
		\end{equation}

		The contributions which are first order in GSR from first-order modefunction corrections have the sources
		\begin{equation}
		S'_{42} = S'_{43} = S'_{44} = S'_{4B} = g S_4,
		\end{equation}
		the windows
		\begin{equation}
		W_{42} = \cos{x}, \quad
		W_{43} = 2 \frac{\sin{x}}{x} - \cos{x}, \quad
		W_{44} = 12 \left(\frac{\sin{x}}{x^3} - \frac{\cos{x}}{x^2} - \frac{\sin{x}}{4x}\right), \quad
		W_{4B} = W_{44},
		\end{equation}
		and the $k$-weights 
		\begin{align*}
			T_{42} =& \frac{1}{2 K} \frac{1}{k_1 k_2 k_3} \Bigl[(\mathbf{k_1}\cdot\mathbf{k_2}) k_3^2 + \text{perm.} \Bigr], \\
			T_{43} =& \frac{1}{2 K} \frac{1}{(K
			 - 2 k_1) (K
			 - 2 k_2) (K
			 - 2 k_3)}\frac{1}{k_1 k_2 k_3} \Bigl[(\mathbf{k_1}\cdot\mathbf{k_2}) k_3^2  \Bigl( 5 k_1^3 - 5 k_1^2 k_2 - 5 k_1 k_2^2 
			 + 5 k_2^3 
			 - 3 k_1^2 k_3 
			 \\&+ 6 k_1 k_2 k_3 - 3 k_2^2 k_3 - k_1 k_3^2 - k_2 k_3^2 - k_3^3 \Bigr)+ \text{perm.}\Bigr], \\ 
			T_{44} =& \frac{1}{3 K}\frac{1}{(K- 2 k_1)^2(K- 2 k_2)^2 (K
			 - 2 k_3)^2} \frac{1}{k_1 k_2 k_3} \Bigl[ (\mathbf{k_1}\cdot\mathbf{k_2}) k_3^2 \Bigl(9 k_1^6 - 2 k_1^5 (9 k_2 + 8 k_3) 
			 \\&+ k_1^4 (-9 k_2^2 + 16 k_2 k_3 - 3 k_3^2) + (k_2 - k_3)^3 (9 k_2^3 + 11 k_2^2 k_3 + 3 k_2 k_3^2 + k_3^3) 
			 \\&+ 4 k_1^3 (9 k_2^3 - k_2 k_3^2 + 4 k_3^3)
 			 + k_1^2 (-9 k_2^4 + 14 k_2^2 k_3^2 - 5 k_3^4) - 
 			 2 k_1 (9 k_2^5 - 8 k_2^4 k_3 + 2 k_2^3 k_3^2 - 3 k_2 k_3^4) \Bigr)
	     	\\&+ \text{perm.} \Bigr], \\ 
			T_{4B} =& -\frac{1}{6}\frac{1}{(K
			 - 2 k_3)^2}\frac{1}{ k_1 k_2  k_3 } \Bigl[2 (\mathbf{k_1}\cdot\mathbf{k_3}) k_2^2 (
			 3 k_1
			 + 2 k_2
			 - 3 k_3)
			 + (\mathbf{k_1}\cdot\mathbf{k_2}) k_3^2 (
			 3 K
			 - 5 k_3)\Bigr] + \Bigl[1 \leftrightarrow 2\Bigr].
		\numberthis
		\end{align*}
		The $k$-weights satisfy the relation
		\begin{equation}
		\sum_{j=2}^{4} T_{4j} + \left[T_{4B} + \text{perm.}\right] = 0,
		\end{equation}
		which ensures that the outside-the-horizon boundary term in $I_{4j}$ for these contributions cancel and therefore that $g S_4$ does not need to be integrated. 

		In the squeezed-limit $x_L / x_S \gg 1$, $x_S \ll 1$, this operator satisfies
		\begin{equation}
		\sum_{j} T_{4j} I_{4j} + \left[T_{4B} I_{4B}(2 k_3) + \text{perm.}\right] = 0,
		\end{equation}
		and therefore this operator has no squeezed contribution.

	\subsection*{\texorpdfstring{$\bm{i=5}$}{i=5}: \texorpdfstring{$\pa^2 \zeta (\pa \chi)^2$}{}}
		
		This operator corresponds to the interaction Hamiltonian
		\begin{equation}
		H_I^{i=5} =  -\int \diffcubed{x} \ a^3 F_{5} \pa^2 \zeta (\pa \chi)^2.
		\end{equation}

		This operator is slow-roll suppressed in GLPV and Horndeski models and even more so in $k$-inflation models.  In the general EFT, it is not
		necessarily suppressed and therefore we compute it to first order in GSR modefunctions. 
		We define the source
		\begin{equation}
		S_5 \equiv \frac{1}{Q}\frac{F_5}{f}.
		\end{equation}
		The contributions which are first order in GSR due to the time variation of the source have the sources
		\begin{equation}
		S_{50} = S_{51} = S_5,
		\end{equation}
		the windows
		\begin{equation}
		W_{50} = x \sin{x}, \quad W_{51} = \cos{x},
		\end{equation}
		and the $k$-weights
		\begin{equation}
		T_{50} = \frac{1}{K^2} \frac{1}{k_1 k_2 k_3} \Bigl[(\mathbf{k_1}\cdot \mathbf{k_2}) k_3^3 +\text{perm.}\Bigr], \quad
		T_{51} = \frac{1}{K^2} \frac{1}{k_1 k_2 k_3} \Bigl[(\mathbf{k_1}\cdot \mathbf{k_2}) k_3^2 (K + k_3)+ \text{perm.}\Bigr].
		\end{equation}

		The contributions which are first order in GSR from first-order modefunction corrections have the sources
		\begin{equation}
		S'_{52} = S'_{53} = S'_{54} =  S'_{5B} = g S_5,
		\end{equation}
		the windows
		\begin{equation}
		W_{52} = \cos{x}, \quad
		W_{53} = 2 \frac{\sin{x}}{x} - \cos{x}, \quad
		W_{54} = 12 \left(\frac{\sin{x}}{x^3} - \frac{\cos{x}}{x^2} - \frac{\sin{x}}{4x}\right), \quad
		W_{5B} = W_{54},
		\end{equation}
		and the $k$-weights 
		\begin{align*}
			T_{52} =& -\frac{1}{4 K} \frac{1}{k_1 k_2 k_3} \Bigl[(\mathbf{k_1}\cdot\mathbf{k_2}) k_3^2 + \text{perm.} \Bigr], \\
			T_{53} =& \frac{1}{4 K} \frac{1}{(K
			 - 2 k_1) (K
			 - 2 k_2) (K
			 - 2 k_3)}\frac{1}{k_1 k_2 k_3} \Bigl[(\mathbf{k_1}\cdot\mathbf{k_2}) k_3^2 \Bigl(k_1^3
			 - k_1^2 k_2
			 - k_1 k_2^2
			 + k_2^3
			 + 3 k_1^2 k_3
			 \\&- 6 k_1 k_2 k_3
			 + 3 k_2^2 k_3
			 + 7 k_1 k_3^2
			 + 7 k_2 k_3^2
			 - 11 k_3^3\Bigr)+ \text{perm.} \Bigr], \\ 
			T_{54} =& \frac{1}{6 K}\frac{1}{(K- 2 k_1)^2(K- 2 k_2)^2 (K
			  - 2 k_3)^2} \frac{1}{k_1 k_2 k_3} \Bigl[ (\mathbf{k_1}\cdot\mathbf{k_2}) k_3^2 \Bigl(k_1^6 - 2 k_1^5 (k_2 - k_3) 
			  \\&-k_1^4 (k_2^2 + 6 k_2 k_3 - 11 k_3^2) + 
	  		 4 k_1^3 (k_2^3 + k_2^2 k_3 - k_2 k_3^2 - 9 k_3^3)
	  		 \\&+ (k_2 - 
	         k_3)^3 (k_2^3 + 5 k_2^2 k_3 + 23 k_2 k_3^2 + 19 k_3^3) + 
	         k_1^2 (-k_2^4 + 4 k_2^3 k_3 - 14 k_2^2 k_3^2 + 
	         4 k_2 k_3^3 + 7 k_3^4) 
	         \\&- 2 k_1 (k_2^5 + 3 k_2^4 k_3 + 2 k_2^3 k_3^2 - 
	         2 k_2^2 k_3^3 + 13 k_2 k_3^4 - 17 k_3^5)\Bigr)
	     	+ \text{perm.} \Bigr], \\ 
			T_{5B} =& \frac{1}{6}\frac{1}{(K
			 - 2 k_3)^2}\frac{1}{ k_1 k_2  k_3 } \Bigl[2 (\mathbf{k_1}\cdot\mathbf{k_3}) k_2^2 (
			 k_1
			 + 2 k_2
			 - k_3)
			 + (\mathbf{k_1}\cdot\mathbf{k_2}) k_3^2 (
			 K
			 - 3 k_3)\Bigr] + \Bigl[ 1 \leftrightarrow 2\Bigr].
		\numberthis
		\end{align*}
		The $k$-weights  obey
		\begin{equation}
		\sum_{j=2}^{4} T_{5j} + \left[T_{5B} + \text{perm.}\right] = 0,
		\end{equation}
		which ensures that $g S_5$ does not need to be integrated.
		
		In the squeezed-limit $x_L / x_S \gg 1$, $x_S \ll 1$, this operator satisfies
		\begin{equation}
		\sum_{j} T_{5j} I_{5j} + \left[T_{5B} I_{5B}(2 k_3) + \text{perm.}\right] = 0,
		\end{equation}
		and therefore this operator has no squeezed contribution.

	\subsection*{\texorpdfstring{$\bm{i=6}$}{i=6}: \texorpdfstring{$\dot{\zeta} \pa^2 \zeta \pa^2 \zeta$}{}}
		
		This operator corresponds to the interaction Hamiltonian
		\begin{equation}
		H_I^{i=6} =  -\int \diffcubed{x} \frac{F_{6}}{{H^3} a}  \dot{\zeta} \pa^2 \zeta \pa^2 \zeta.
		\end{equation}

		This operator, and all subsequent operators, is not present in the Horndeski or beyond-Horndeski GLPV class. 
		We define the source
		\begin{equation}
		S_6 \equiv \frac{1}{Q c_s^4} \left(\frac{c_s}{a H s}\right)^{3}\frac{F_6}{f}.
		\end{equation}

		The initial windows in this case have high powers of $x$ and are therefore challenging to integrate numerically, so we follow the approach used by Ref.~\cite{Adshead:2013zfa} in the $i=3$ case and integrate them by parts, at the expense of placing additional derivatives on the sources, such that we have the sources

		\begin{equation}
		S_{60} = S_6''', \quad
		S_{61} = S_6'', \quad
		S_{62} = S_6', \quad 
		S_{63} = S_6,
		\end{equation}
		the windows
		\begin{equation}
		W_{60} = W_{61} = W_{62} = W_{63} = x \sin{x} + \cos{x},
		\end{equation}
		and the $k$-weights 
		\begin{align*}
		T_{60} &= \frac{k_1 k_2 k_3}{K^5} \left[\sum_{i>j} k_i k_j \right],\quad
		&&T_{61} = \frac{k_1 k_2 k_3}{K^5} \left[2 K^2 + 9\sum_{i>j} k_i k_j \right],\\
		T_{62} &= \frac{13 k_1 k_2 k_3}{K^5} \left[K^2 + 2\sum_{i>j} k_i k_j \right],
		&&T_{63} = \frac{6 k_1 k_2 k_3}{K^5} \left[3 K^2 + 4\sum_{i>j} k_i k_j \right].
	    \numberthis
		\end{align*}

		The contributions which are first order in GSR from first-order modefunction corrections have the sources

		\begin{equation}
		S'_{64} = S'_{65} = S'_{66} = S'_{67} = S'_{68} = S'_{6B} = g S_6,
		\end{equation}

		the windows
		\begin{align*}
		&W_{64} = x^3 \sin{x}, \quad
		&&W_{65} = x^2 \cos{x}, \\
		&W_{66} = \cos{x} + x \sin{x},
		&&W_{67} = 2 \frac{\sin{x}}{x} - \cos{x}, \\
		&W_{68} = 12 \left(\frac{\sin{x}}{x^3} - \frac{\cos{x}}{x^2} - \frac{\sin{x}}{4x}\right), 
		&&W_{6B} = W_{68},
		\numberthis
		\end{align*}
		and the $k$-weights 
		\begin{align*}
			T_{64} =& -\frac{1}{2 K^4}\frac{k_1 k_2 k_3}{(K-2 k_1)(K- 2k_2) (K-2k_3)} \sum_{i\neq j \neq h} \Bigl[ 2 k_i^3 k_j -2 k_i^2 k_h^2 + k_i^2 k_j k_h \Bigr],\\
			T_{65} =& \frac{1}{K^4}\frac{k_1 k_2 k_3}{(K
			 - 2 k_1)^2 (K
			 - 2 k_2)^2 (K
			 - 2 k_3)^2} \sum_{i\neq j\neq h} \Bigl[ k_i^7
			 + 2 k_i^6 k_j
			 - 50 k_i^5 k_j^2 
			 + 46 k_i^4 k_j^3
			 \\&+ 8 k_i^5 k_j k_h
			 + 14 k_i^4 k_j^2 k_h
			 - 32 k_i^3 k_j^3 k_h
			 + 2 k_i^3 k_j^2 k_h^2  \Bigr],\\
			T_{66} =& \frac{1}{K^4}\frac{k_1 k_2 k_3}{(K
			 - 2 k_1)^3 (K
			 - 2 k_2)^3 (K
			 - 2 k_3)^3} \sum_{i\neq j\neq h} \Bigl[ \frac{3}{2} k_i^{10} + 26 k_i^9 k_j - 361 k_i^8 k_j^2 + 
			 600 k_i^7 k_j^3 \\
			 &+ 358 k_i^6 k_j^4 
			 - 626 k_i^5 k_j^5 + 
			 77 k_i^8 k_j k_h + 344 k_i^7 k_j^2 k_h - 
			 1064 k_i^6 k_j^3 k_h + 540 k_i^5 k_j^4 k_h \\
			 &+ 170 k_i^6 k_j^2 k_h^2 + 168 k_i^5 k_j^3 k_h^2 - 
			 491 k_i^4 k_j^4 k_h^2 + 340 k_i^4 k_j^3 k_h^3 \Bigr], \\
			T_{67} =& \frac{1}{K^4} \frac{3 k_1 k_2 k_3}{(K
			 - 2 k_1)^4 (K
			 - 2 k_2)^4 (K
			 - 2 k_3)^4} \sum_{i\neq j\neq h} \Bigl[ \frac{11}{2} k_i^{13} + 47 k_i^{12} k_j - 646 k_i^{11} k_j^2 \\
			 &+   1130 k_i^{10} k_j^3 + 601 k_i^9 k_j^4 - 3059 k_i^8 k_j^5 + 1916 k_i^7 k_j^6 + 54 k_i^{11} k_j k_h + 686 k_i^{10} k_j^2 k_h \\
  			 &- 2884 k_i^9 k_j^3 k_h + 1953 k_i^8 k_j^4 k_h + 2776 k_i^7 k_j^5 k_h - 2686 k_i^6 k_j^6 k_h + 51 k_i^9 k_j^2 k_h^2 + 418 k_i^8 k_j^3 k_h^2 \\
  			 &-  5180 k_i^7 k_j^4 k_h^2 + 
  			 4620 k_i^6 k_j^5 k_h^2 + 1832 k_i^7 k_j^3 k_h^3 + 
  			 500 k_i^6 k_j^4 k_h^3 - 2828 k_i^5 k_j^5 k_h^3 + 
  			 551 k_i^5 k_j^4 k_h^4  \Bigr],\\
  			T_{68} =& \frac{1}{K^4}\frac{4 k_1 k_2 k_3}{(K
			 - 2 k_1)^5 (K
			 - 2 k_2)^5 (K
			 - 2 k_3)^5} \sum_{i\neq j\neq h} \Bigl[ \frac{9}{2} k_i^{16} + 18 k_i^{15} k_j - 436 k_i^{14} k_j^2 + 1030 k_i^{13} k_j^3 \\
			 &+ 196 k_i^{12} k_j^4 - 3198 k_i^{11} k_j^5 + 
			 2996 k_i^{10} k_j^6 + 2150 k_i^9 k_j^7 - 2765 k_i^8 k_j^8 + 9 k_i^{14} k_j k_h + 790 k_i^{13} k_j^2 k_h \\
			 &- 2890 k_i^{12} k_j^3 k_h + 2802 k_i^{11} k_j^4 k_h + 3282 k_i^{10} k_j^5 k_h - 10490 k_i^9 k_j^6 k_h + 6470 k_i^8 k_j^7 k_h - 114 k_i^{12} k_j^2 k_h^2 \\ 
			 &+ 1484 k_i^{11} k_j^3 k_h^2 - 7452 k_i^{10} k_j^4 k_h^2 + 8458 k_i^9 k_j^5 k_h^2 + 8116 k_i^8 k_j^6 k_h^2 - 10732 k_i^7 k_j^7 k_h^2 + 2502 k_i^{10} k_j^3 k_h^3 \\&+  970 k_i^9 k_j^4 k_h^3 - 15814 k_i^8 k_j^5 k_h^3 + 10216 k_i^7 k_j^6 k_h^3 + 4710 k_i^8 k_j^4 k_h^4 + 836 k_i^7 k_j^5 k_h^4 - 6772 k_i^6 k_j^6 k_h^4 \\&+ 3730 k_i^6 k_j^5 k_h^5  \Bigr], \\
			 T_{6B} &= \frac{2 k_1 k_2 k_3}{(K-2k_3)^5} \Bigl[3 (k_1^2 + k_2^2 + k_3^2)+ 10 (k_1 k_2 - k_2 k_3 - k_1 k_3) \Bigr].
		\numberthis
		\end{align*}

		The sums here are over the 6 permutations of $k_1, k_2, k_3$. The $k$-weights  obey
		\begin{equation}
		\sum_{j=6}^{8} T_{6j} + (T_{6B} + \text{perm.}) = 0,
		\end{equation}
		which ensures we do not have to integrate $g S_6$.

		In the squeezed-limit $x_L / x_S \gg 1$, $x_S \ll 1$, this operator satisfies
		\begin{equation}
		\sum_{j} T_{6j} I_{6j} + \left(T_{6B} I_{6B}(2 k_3) + \text{perm.}\right) = 0,
		\end{equation}
		and therefore this operator has no squeezed contribution. We do not integrate by parts the windows $W_{64}$ and $W_{65}$ because this would obscure the above cancellation.

	\subsection*{\texorpdfstring{$\bm{i=7}$}{i=7}: \texorpdfstring{$(\pa_a \pa_b \zeta)^2 \pa^2 \zeta$}{}}

		This operator corresponds to the interaction Hamiltonian
		\begin{equation}
		H_I^{i=7} =  -\int \diffcubed{x} \frac{F_{7}}{{H^4} a^3} \pa^2 \zeta (\pa_a \pa_b \zeta)^2.
		\end{equation}

		We define the source
		\begin{equation}
		S_7 \equiv{ \frac{1}{Q c_s^6}  \left( \frac{c_s}{a Hs} \right)^4} \frac{F_7}{f}.
		\end{equation}

		The contributions from the variation of the source again require significant integration by parts. The sources become

		\begin{equation}
		S_{70} = S_7'''',\quad
		S_{71} = S_7''',\quad
		S_{72} = S_7'',\quad
		S_{73} = S_7',\quad
		S_{74} = S_7,
		\end{equation}
		with the windows
		\begin{equation}
		W_{70} = W_{71} = W_{72} = W_{73} = W_{74} = x \sin{x} + \cos{x},
		\end{equation}
		and the $k$-weights 
		\begin{align*}
		T_{70} &= \frac{T_7}{K^6} \Bigl[k_1 k_2 k_3\Bigr],\\
		T_{71} &= \frac{T_7}{K^6} \Bigl[14 k_1 k_2 k_3 + K \sum_{i>j} k_i k_j \Bigr],\\
		T_{72} &= \frac{T_7}{K^6} \Bigl[K^3 + 71 k_1 k_2 k_3 + 9 K \sum_{i>j} k_i k_j \Bigr],\\
		T_{73} &= 2 \frac{T_7}{K^6} \Bigl[3 K^3 + 77 k_1 k_2 k_3 + 13 K \sum_{i>j} k_i k_j \Bigr],\\
		T_{74} &= 8 \frac{T_7}{K^6} \Bigl[K^3 + 15 k_1 k_2 k_3 + 3 K \sum_{i>j} k_i k_j \Bigr],
	    \numberthis
		\end{align*}
		wherein
		\begin{equation}
		T_7 \equiv\frac{1}{k_1 k_2 k_3} \Bigl[(\mathbf{k_1}\cdot\mathbf{k_2})^2 k_3^2 + \text{perm.}\Bigr].
		\end{equation}

		The contributions which are first order in GSR from first-order modefunction corrections have the sources
		\begin{equation}
		S'_{75} = S'_{76} = S'_{77} = S'_{78} = S'_{79} = S'_{7a} = S'_{7B} = g S_7,
		\end{equation}

		the windows
		\begin{align*}
		&W_{75} = x^4 \cos{x}, \quad
		W_{76} = x^3 \sin{x}, \quad
		W_{77} = x^2 \cos{x}, \quad
		W_{78} = \cos{x} + x \sin{x}, \quad\\
		&W_{79} = 2 \frac{\sin{x}}{x} - \cos{x}, \quad
		W_{7a} = 12 \left(\frac{\sin{x}}{x^3} - \frac{\cos{x}}{x^2} - \frac{\sin{x}}{4x}\right), \quad
		W_{7B} = W_{7a},
		\numberthis
		\end{align*}
		and the $k$-weights 
		\begin{align*}
			T_{75} &= \frac{-1}{2 K^5} \frac{T_7 k_1 k_2 k_3}{(K-2 k_1)(K-2 k_2)(K-2 k_3)} \sum_{i\neq j\neq k} \Bigl[ k_i^2 - 2 k_i k_j \Bigr], \\
			T_{76} &= \frac{-1}{2 K^5} \frac{T_7}{(K-2 k_1)^2(K-2 k_2)^2(K-2 k_3)^2} \sum_{i\neq j\neq k}  \Bigl[ 2 k_i^7 k_j
			 - 4 k_i^6 k_j^2
			 - 2 k_i^5 k_j^3
			 + 4 k_i^4 k_j^4
			 \\&+ 17 k_i^6 k_j k_h
			 - 50 k_i^5 k_j^2 k_h
			 + 14 k_i^4 k_j^3 k_h
			 + 6 k_i^4 k_j^2 k_h^2
			 + 42 k_i^3 k_j^3 k_h^2
			 \Bigr],\\
			T_{77} &= \frac{1}{K^5}\frac{T_7}{(K
			 - 2 k_1)^3 (K
			 - 2 k_2)^3 (K
			 - 2 k_3)^3} \sum_{i\neq j\neq h} \Bigl[ \frac{k_i^{11}}{2}
			 + 17 k_i^{10} k_j
			 - 27 k_i^9 k_j^2
			 - 27 k_i^8 k_j^3
			 \\&+ 58 k_i^7 k_j^4
			 - 22 k_i^6 k_j^5
			 + 42 k_i^9 k_j k_h
			 - 299 k_i^8 k_j^2 k_h
			 + 160 k_i^7 k_j^3 k_h
			 + 282 k_i^6 k_j^4 k_h
			 \\&- 244 k_i^5 k_j^5 k_h
			 - 30 k_i^7 k_j^2 k_h^2
			 + 324 k_i^6 k_j^3 k_h^2
			 + 62 k_i^5 k_j^4 k_h^2
			 + 64 k_i^5 k_j^3 k_h^3
			 - 585 k_i^4 k_j^4 k_h^3 \Bigr], \\
			T_{78} &= \frac{1}{K^5}\frac{T_7}{(K
			 - 2 k_1)^4 (K
			 - 2 k_2)^4 (K
			 - 2 k_3)^4} \sum_{i\neq j\neq h} \Bigl[ \frac{17}{2} k_i^{14}
			 + 84 k_i^{13} k_j
			 - 333 k_i^{12} k_j^2
			 - 8 k_i^{11} k_j^3
			 \\&+ 897 k_i^{10} k_j^4
			 - 724 k_i^9 k_j^5
			 - 581 k_i^8 k_j^6
			 + 648 k_i^7 k_j^7
			 + 138 k_i^{12} k_j k_h
			 - 1592 k_i^{11} k_j^2 k_h
			 + 1992 k_i^{10} k_j^3 k_h
			 \\&+ 1452 k_i^9 k_j^4 k_h
			 - 6292 k_i^8 k_j^5 k_h
			 + 4080 k_i^7 k_j^6 k_h
			 + 791 k_i^{10} k_j^2 k_h^2
			 + 4072 k_i^9 k_j^3 k_h^2
			 - 1923 k_i^8 k_j^4 k_h^2
			 \\&- 2480 k_i^7 k_j^5 k_h^2
			 + 674 k_i^6 k_j^6 k_h^2
			 + 2796 k_i^8 k_j^3 k_h^3
			 - 11856 k_i^7 k_j^4 k_h^3
			 + 208 k_i^6 k_j^5 k_h^3
			 + 1089 k_i^6 k_j^4 k_h^4
			 \\&+ 9252 k_i^5 k_j^5 k_h^4\Bigr],\\
			 T_{79} &= \frac{1}{K^5}\frac{T_7}{(K
			 - 2 k_1)^5 (K
			 - 2 k_2)^5 (K
			 - 2 k_3)^5} \sum_{i\neq j\neq h} \Bigl[ \frac{67}{2} k_i^{17}
			 + 209 k_i^{16} k_j
			 - 1326 k_i^{15} k_j^2
			 + 1182 k_i^{14} k_j^3
			 \\&+ 3398 k_i^{13} k_j^4
			 - 8054 k_i^{12} k_j^5
			 + 346 k_i^{11} k_j^6
			 + 15158 k_i^{10} k_j^7
			 - 10980 k_i^9 k_j^8
			 + 213 k_i^{15} k_j k_h
			 - 4276 k_i^{14} k_j^2 k_h
			 \\&+ 7950 k_i^{13} k_j^3 k_h
			 + 1316 k_i^{12} k_j^4 k_h
			 - 26406 k_i^{11} k_j^5 k_h
			 + 29876 k_i^{10} k_j^6 k_h
			 + 18030 k_i^9 k_j^7 k_h
			 - 27125 k_i^8 k_j^8 k_h
			 \\&+ 1804 k_i^{13} k_j^2 k_h^2
			 + 13074 k_i^{12} k_j^3 k_h^2
			 - 21330 k_i^{11} k_j^4 k_h^2
			 + 4896 k_i^{10} k_j^5 k_h^2
			 + 55356 k_i^9 k_j^6 k_h^2
			 - 50002 k_i^8 k_j^7 k_h^2
			 \\&- 306 k_i^{11} k_j^3 k_h^3
			 - 94698 k_i^{10} k_j^4 k_h^3
			 + 27666 k_i^9 k_j^5 k_h^3
			 + 80442 k_i^8 k_j^6 k_h^3
			 - 35004 k_i^7 k_j^7 k_h^3
			 - 17880 k_i^9 k_j^4 k_h^4
			 \\&+ 121830 k_i^8 k_j^5 k_h^4
			 + 25244 k_i^7 k_j^6 k_h^4
			 + 6282 k_i^7 k_j^5 k_h^5
			 - 132496 k_i^6 k_j^6 k_h^5 \Bigr],\\
			 T_{7a} &= \frac{2}{3 K^3}\frac{T_7}{(K
			 - 2 k_1)^6 (K
			 - 2 k_2)^6 (K
			 - 2 k_3)^6} \sum_{i\neq j\neq h} \Bigl[ \frac{71}{2} k_i^{18}
			 - 1617 k_i^{16} k_j^2
			 + 5600 k_i^{15} k_j^3
			 - 6420 k_i^{14} k_j^4
			 \\&- 6720 k_i^{13} k_j^5
			 + 28252 k_i^{12} k_j^6
			 - 23520 k_i^{11} k_j^7
			 - 20286 k_i^{10} k_j^8
			 + 24640 k_i^9 k_j^9
			 + 4788 k_i^{14} k_j^2 k_h^2 
			 - 6720 k_i^{13} k_j^3 k_h^2
			 \\&- 27804 k_i^{12} k_j^4 k_h^2
			 + 84672 k_i^{11} k_j^5 k_h^2
			 - 78120 k_i^{10} k_j^6 k_h^2
			 - 77952 k_i^9 k_j^7 k_h^2
			 + 97965 k_i^8 k_j^8 k_h^2
			 + 2800 k_i^{12} k_j^3 k_h^3
			 \\&- 23520 k_i^{11} k_j^4 k_h^3
			 - 6720 k_i^{10} k_j^5 k_h^3
			 + 49280 k_i^9 k_j^6 k_h^3
			 - 23520 k_i^8 k_j^7 k_h^3
			 + 105126 k_i^{10} k_j^4 k_h^4
			 - 77952 k_i^9 k_j^5 k_h^4
			 \\&- 194460 k_i^8 k_j^6 k_h^4
			 + 119904 k_i^7 k_j^7 k_h^4
			 + 42336 k_i^8 k_j^5 k_h^5
			 - 77952 k_i^7 k_j^6 k_h^5
			 + 91000 k_i^6 k_j^6 k_h^6 \Bigr],\\
			 T_{7B} &= \frac{8 T_7}{3 (K
			 - 2 k_3)^6} \Bigl[ k_1^3 + k_2^3 - k_3^3 + 6 \left( k_1^2 k_2 + k_1 k_2^2 - k_1^2 k_3 - k_2^2 k_3 +k_1 k_3^2 + k_2 k_3^2 \right) - 30 k_1 k_2 k_3\Bigr].
			\numberthis
		\end{align*}

		These $k$-weights  obey:
		\begin{equation}
		\sum_{j=8}^{a} T_{7j} + [T_{7B} + \text{perm.}] = 0,
		\end{equation}
		which ensures we do not have to integrate $g S_7$.

		In the squeezed-limit $x_L / x_S \gg 1$, $x_S \ll 1$, this operator satisfies
		\begin{equation}
		\sum_{j} T_{7j} I_{7j} + \left[T_{7B} I_{7B}(2 k_3) + \text{perm.}\right] = 0,
		\end{equation}
		and therefore this operator has no squeezed contribution. We do not integrate by parts the windows $W_{75}$, $W_{76}$, and $W_{77}$ because this would obscure the above cancellation.

	\subsection*{\texorpdfstring{$\bm{i=8}$}{i=8}: \texorpdfstring{$(\pa^2 \zeta) (\pa^2 \zeta) (\pa^2 \zeta)$}{}}

		This operator corresponds to the interaction Hamiltonian
		\begin{equation}
		H_I^{i=8} =  -\int \diffcubed{x} \frac{F_{8}}{{H^4} a^3} (\pa^2 \zeta)^3.
		\end{equation}
		
		We define the source
		\begin{equation}
		S_8 \equiv{ \frac{1}{Q c_s^6}  \left( \frac{c_s}{a Hs} \right)^4} \frac{F_8}{f}.
		\end{equation}

		The contribution from this operator is equal to the contribution from the $i=7$ operator with the substitutions 
		\begin{equation}
		S_7 \to S_8, \quad T_7 \to T_8 \equiv 3 k_1 k_2 k_3.
		\end{equation}

		Again, in the squeezed-limit this operator satisfies 
		\begin{equation}
		\sum_{j} T_{8j} I_{8j} + \left[T_{8B} I_{8B}(2 k_3) + \text{perm.}\right] = 0,
		\end{equation}
		and therefore this operator has no squeezed contribution.

	\subsection*{\texorpdfstring{$\bm{i=9}$}{i=9}: \texorpdfstring{$(\pa^2 \zeta) (\pa_a \pa_b \zeta) (\pa_a \pa_b \chi)$}{}}

		The final operator in the cubic action of the unified EFT of inflation corresponds to the interaction Hamiltonian
		\begin{equation}
		H_I^{i=9} =  -\int \diffcubed{x} \frac{F_{9}}{{H^3} a} (\pa^2 \zeta) (\pa_a \pa_b \zeta)  (\pa_a \pa_b \chi).
		\end{equation}

		We define the source
		\begin{equation}
		S_9 \equiv{ \frac{1}{Q c_s^4}  \left( \frac{c_s}{a Hs} \right)^3} \frac{F_9}{f}.
		\end{equation}

		The contributions from the variation of the source once again require significant integration by parts, and the resulting sources are
		\begin{equation}
		S_{90} = S_9''',\quad
		S_{91} = S_9'',\quad
		S_{92} = S_9', \quad
		S_{93} = S_9.
		\end{equation}
		The corresponding windows and $k$-weights  are
		\begin{equation}
		W_{90} = W_{91} = W_{92} = W_{93} = x \sin{x} + \cos{x},
		\numberthis
		\end{equation}
		\begin{align*}
		T_{90} &= \frac{1}{2 K^5}\frac{1}{k_1 k_2 k_3} \Bigl[(\mathbf{k_1}\cdot\mathbf{k_2})^2 k_3^3 (k_1+k_2) + \text{perm.}\Bigr],\\
		T_{91} &= \frac{1}{2 K^5}\frac{1}{k_1 k_2 k_3}  \Bigl[(\mathbf{k_1}\cdot\mathbf{k_2})^2 k_3^2 (9 (k_1+k_2) k_3 + K (K+k_3)) + \text{perm.}\Bigr], \\
		T_{92} &=  \frac{1}{2 K^5}\frac{1}{k_1 k_2 k_3}  \Bigl[(\mathbf{k_1}\cdot\mathbf{k_2})^2 k_3^2 (26 (k_1+k_2) k_3 + 5 K (K+k_3) + 2 K^2) + \text{perm.}\Bigr], \\
		T_{93} &= \frac{1}{K^5}\frac{1}{k_1 k_2 k_3}  \Bigl[(\mathbf{k_1}\cdot\mathbf{k_2})^2 k_3^2 (12 (k_1+k_2) k_3 + 3 K (K+k_3) + 2 K^2) + \text{perm.}\Bigr].
	    \numberthis
		\end{align*}

		The contributions which are first order in GSR from first-order modefunction corrections have the sources, windows, and weights
		\begin{equation}
		S'_{94} = S'_{95} = S'_{96} = S'_{97} = S'_{98} = S'_{9B} = g S_9,
		\end{equation}
		\begin{align*}
		&W_{94} = x^3 \sin{x}, \quad
		&&W_{95} = x^2 \cos{x}, \\
		&W_{96} = \cos{x} + x \sin{x},
		&&W_{97} = 2 \frac{\sin{x}}{x} - \cos{x}, \\
		&W_{98} = 12 \left(\frac{\sin{x}}{x^3} - \frac{\cos{x}}{x^2} - \frac{\sin{x}}{4x}\right), 
		&&W_{9B} = W_{98},
		\numberthis
		\end{align*}
		\begin{align*}
			T_{94} &= \frac{-1}{8 k_1 k_2 k_3} \frac{1}{K^4 (K- 2 k_1) (K - 2 k_2) (K- 2 k_3)}  \sum_{i\neq j\neq h} \Bigl[ k_i^9 k_j
			 - 2 k_i^8 k_j^2
			 + 2 k_i^6 k_j^4
			 - k_i^5 k_j^5
			 \\&+ 2 k_i^8 k_j k_h
			 - 3 k_i^7 k_j^2 k_h
			 - 9 k_i^6 k_j^3 k_h
			 + 7 k_i^5 k_j^4 k_h
			 + 4 k_i^6 k_j^2 k_h^2
			 + 3 k_i^5 k_j^3 k_h^2
			 - 6 k_i^4 k_j^4 k_h^2
			 + 3 k_i^4 k_j^3 k_h^3 \Bigr] \\
		 	T_{95} &= \frac{1}{8 k_1 k_2 k_3} \frac{1}{K^4 (K- 2 k_1)^2 (K - 2 k_2)^2 (K- 2 k_3)^2} \sum_{i\neq j\neq h} \Bigl[  k_i^{13}
			 + 5 k_i^{12} k_j
			 - 55 k_i^{11} k_j^2
			 + 29 k_i^{10} k_j^3
			 \\&+ 109 k_i^9 k_j^4
			 - 68 k_i^8 k_j^5
			 - 22 k_i^7 k_j^6
			 + 16 k_i^{11} k_j k_h
			 - 38 k_i^{10} k_j^2 k_h
			 - 128 k_i^9 k_j^3 k_h
			 + 171 k_i^8 k_j^4 k_h
			 \\&+ 96 k_i^7 k_j^5 k_h
			 - 138 k_i^6 k_j^6 k_h
			 + 119 k_i^9 k_j^2 k_h^2
			 - 23 k_i^8 k_j^3 k_h^2
			 - 410 k_i^7 k_j^4 k_h^2
			 + 288 k_i^6 k_j^5 k_h^2
			 \\&+ 160 k_i^7 k_j^3 k_h^3
			 - 6 k_i^6 k_j^4 k_h^3
			 - 192 k_i^5 k_j^5 k_h^3
			 + 68 k_i^5 k_j^4 k_h^4  \Bigr] \\
			T_{96} =& \frac{1}{4 K^4} \frac{1}{(K- 2 k_1)^3 (K - 2 k_2)^3 (K- 2 k_3)^3} \frac{1}{k_1 k_2 k_3} \sum_{i\neq j\neq h} \Bigl[ 
			\frac{3}{2} k_i^{16}
			 + 21 k_i^{15} k_j
			 - 210 k_i^{14} k_j^2
			 + 247 k_i^{13} k_j^3
			 \\&+ 496 k_i^{12} k_j^4
			 - 867 k_i^{11} k_j^5
			 - 142 k_i^{10} k_j^6
			 + 599 k_i^9 k_j^7
			 - 147 k_i^8 k_j^8
			 + 52 k_i^{14} k_j k_h
			 + 6 k_i^{13} k_j^2 k_h
			 \\&- 631 k_i^{12} k_j^3 k_h
			 + 703 k_i^{11} k_j^4 k_h
			 + 422 k_i^{10} k_j^5 k_h
			 - 2036 k_i^9 k_j^6 k_h
			 + 1411 k_i^8 k_j^7 k_h
			 + 625 k_i^{12} k_j^2 k_h^2
			 \\&- 596 k_i^{11} k_j^3 k_h^2
			 - 3002 k_i^{10} k_j^4 k_h^2
			 + 3354 k_i^9 k_j^5 k_h^2
			 + 1962 k_i^8 k_j^6 k_h^2
			 - 2764 k_i^7 k_j^7 k_h^2
			 + 1538 k_i^{10} k_j^3 k_h^3
			 \\&- 493 k_i^9 k_j^4 k_h^3
			 - 5203 k_i^8 k_j^5 k_h^3
			 + 3600 k_i^7 k_j^6 k_h^3
			 + 2412 k_i^8 k_j^4 k_h^4
			 - 210 k_i^7 k_j^5 k_h^4
			 - 2318 k_i^6 k_j^6 k_h^4
			 \\&+ 1252 k_i^6 k_j^5 k_h^5 \Bigr],\\
			 T_{97} =& \frac{1}{4 K^4} \frac{1}{(K- 2 k_1)^4 (K - 2 k_2)^4 (K- 2 k_3)^4} \frac{1}{k_1 k_2 k_3} \sum_{i\neq j\neq h} \Bigl[ \frac{23}{2} k_i^{19}
			 + 91 k_i^{18} k_j
			 - 1117 k_i^{17} k_j^2
			 + 1567 k_i^{16} k_j^3
			 \\&+ 2572 k_i^{15} k_j^4
			 - 6532 k_i^{14} k_j^5
			 + 1068 k_i^{13} k_j^6
			 + 6972 k_i^{12} k_j^7
			 - 6526 k_i^{11} k_j^8
			 + 1882 k_i^{10} k_j^9
			 + 102 k_i^{17} k_j k_h
			 \\&+ 515 k_i^{16} k_j^2 k_h
			 - 4224 k_i^{15} k_j^3 k_h
			 + 3740 k_i^{14} k_j^4 k_h
			 + 5904 k_i^{13} k_j^5 k_h
			 - 15220 k_i^{12} k_j^6 k_h
			 + 11136 k_i^{11} k_j^7 k_h
			 \\&+ 10874 k_i^{10} k_j^8 k_h
			 - 13020 k_i^9 k_j^9 k_h
			 + 2744 k_i^{15} k_j^2 k_h^2
			 - 5072 k_i^{14} k_j^3 k_h^2
			 - 19324 k_i^{13} k_j^4 k_h^2
			 + 32708 k_i^{12} k_j^5 k_h^2
			 \\&+ 15632 k_i^{11} k_j^6 k_h^2
			 - 63280 k_i^{10} k_j^7 k_h^2
			 + 34450 k_i^9 k_j^8 k_h^2
			 + 12768 k_i^{13} k_j^3 k_h^3
			 - 7436 k_i^{12} k_j^4 k_h^3
			 - 57408 k_i^{11} k_j^5 k_h^3
			 \\&+ 62608 k_i^{10} k_j^6 k_h^3
			 + 36096 k_i^9 k_j^7 k_h^3
			 - 51667 k_i^8 k_j^8 k_h^3
			 + 29166 k_i^{11} k_j^4 k_h^4
			 - 18996 k_i^{10} k_j^5 k_h^4
			 - 91628 k_i^9 k_j^6 k_h^4
			 \\&+ 72740 k_i^8 k_j^7 k_h^4
			 + 46008 k_i^9 k_j^5 k_h^5
			 - 7180 k_i^8 k_j^6 k_h^5
			 - 40512 k_i^7 k_j^7 k_h^5
			 + 17360 k_i^7 k_j^6 k_h^6 \Bigr],\\
			 T_{98} =& \frac{1}{3 K^4} \frac{1}{(K- 2 k_1)^5 (K - 2 k_2)^5 (K- 2 k_3)^5} \frac{1}{k_1 k_2 k_3} \sum_{i\neq j\neq h} \Bigl[ 
			\frac{17}{2} k_i^{22}
			 + 34 k_i^{21} k_j
			 - 749 k_i^{20} k_j^2
			 + 1548 k_i^{19} k_j^3
			 \\&+ 1407 k_i^{18} k_j^4
			 - 6854 k_i^{17} k_j^5
			 + 3605 k_i^{16} k_j^6
			 + 8464 k_i^{15} k_j^7
			 - 12022 k_i^{14} k_j^8
			 + 2212 k_i^{13} k_j^9
			 + 7742 k_i^{12} k_j^{10}
			 \\&- 5404 k_i^{11} k_j^{11}
			 + 17 k_i^{20} k_j k_h
			 + 988 k_i^{19} k_j^2 k_h
			 - 4132 k_i^{18} k_j^3 k_h
			 + 3706 k_i^{17} k_j^4 k_h
			 + 7546 k_i^{16} k_j^5 k_h
			 \\&- 21936 k_i^{15} k_j^6 k_h
			 + 13904 k_i^{14} k_j^7 k_h
			 + 21092 k_i^{13} k_j^8 k_h
			 - 41628 k_i^{12} k_j^9 k_h
			 + 20392 k_i^{11} k_j^{10} k_h
			 \\&+ 1625 k_i^{18} k_j^2 k_h^2
			 - 3956 k_i^{17} k_j^3 k_h^2
			 - 17609 k_i^{16} k_j^4 k_h^2
			 + 37216 k_i^{15} k_j^5 k_h^2
			 + 13592 k_i^{14} k_j^6 k_h^2
			 \\&- 97792 k_i^{13} k_j^7 k_h^2
			 + 70838 k_i^{12} k_j^8 k_h^2
			 + 63544 k_i^{11} k_j^9 k_h^2
			 - 69322 k_i^{10} k_j^{10} k_h^2
			 + 14222 k_i^{16} k_j^3 k_h^3
			 \\&- 11648 k_i^{15} k_j^4 k_h^3
			 - 80480 k_i^{14} k_j^5 k_h^3
			 + 111680 k_i^{13} k_j^6 k_h^3
			 + 43616 k_i^{12} k_j^7 k_h^3
			 - 228872 k_i^{11} k_j^8 k_h^3
			 \\&+ 143800 k_i^{10} k_j^9 k_h^3
			 + 43694 k_i^{14} k_j^4 k_h^4
			 - 37832 k_i^{13} k_j^5 k_h^4
			 - 184404 k_i^{12} k_j^6 k_h^4
			 + 224512 k_i^{11} k_j^7 k_h^4
			 \\&+ 113218 k_i^{10} k_j^8 k_h^4
			 - 178738 k_i^9 k_j^9 k_h^4
			 + 111452 k_i^{12} k_j^5 k_h^5
			 - 64928 k_i^{11} k_j^6 k_h^5
			 - 290912 k_i^{10} k_j^7 k_h^5
			 \\&+ 213340 k_i^9 k_j^8 k_h^5
			 + 143380 k_i^{10} k_j^6 k_h^6
			 - 24816 k_i^9 k_j^7 k_h^6
			 - 119553 k_i^8 k_j^8 k_h^6
			 + 61512 k_i^8 k_j^7 k_h^7
  			\Bigr],\\
  			T_{9B} &= \frac{1}{3} \frac{1}{(K- 2 k_3)^5} \frac{1}{k_1 k_2 k_3} \Bigl[(\mathbf{k_1}\cdot\mathbf{k_3})^2 k_2^2 (8 k_2 ^2 + 25 k_2 (k_1 - k_3) + 5 (k_1 - k_3)^2)
  			\\&+ \frac{1}{2} (\mathbf{k_1}\cdot\mathbf{k_2})^2 k_3^2 (5 (k_1^2 + k_2^2) + 8 k_3^2 + 10 k_1 k_2 - 25 k_3 (k_1 + k_2)) \Bigr] + \Bigl[1 \leftrightarrow 2 \Bigr].
		\numberthis
		\end{align*}
		
		These $k$-weights  obey
		\begin{equation}
		\sum_{j=6}^{8} T_{9j} + \left[T_{9B} + \text{perm.}\right] = 0,
		\end{equation}
		which ensures we do not have to integrate $g S_9$.

		In the squeezed-limit $x_L / x_S \gg 1$, $x_S \ll 1$, this operator satisfies
		\begin{equation}
		\sum_{j} T_{9j} I_{9j} + \left[T_{9B} I_{9B}(2 k_3) + \text{perm.}\right] = 0,
		\end{equation}
		and therefore this operator has no squeezed contribution. We do not integrate by parts the windows $W_{94}$ and $W_{95}$ because this would obscure the above cancellation.

\bibliographystyle{apsrev4-1}
\bibliography{../../references.bib}

\end{document}